%% file: TTUR.tex
\documentclass[prl,nofootinbib,twocolumn,showkeys,superscriptaddress,preprintnumbers,floatfix]{revtex4-1}

\usepackage{etex}
\usepackage{subfigure}
\usepackage{ifpdf}
\usepackage{hyperref}
\usepackage{dcolumn}
\usepackage{url}
\usepackage{amsmath}
\usepackage{amscd}
\usepackage{amsfonts}
\usepackage{amssymb}
\usepackage{bm}   % bold math
\usepackage{bbm}
\usepackage{verbatim}
\usepackage{comment}
\usepackage{stmaryrd}
\usepackage{amsthm}
\usepackage{xcolor}
\usepackage{setspace}
\usepackage{braket}
\usepackage{empheq}
\usepackage{perpage}
\usepackage[english]{babel}
\newtheorem{theorem}{Theorem}%the perpage package
\MakePerPage{footnote}

\usepackage{tikz}
\usetikzlibrary{arrows}
\usetikzlibrary{automata}
\usetikzlibrary{decorations.pathreplacing}
\usetikzlibrary{positioning}
\usetikzlibrary{plotmarks}
\usetikzlibrary{calc}
\usetikzlibrary{patterns}
\usetikzlibrary{external}
\usepackage{pgfplots}
\pgfplotsset{compat=newest}

\usepackage{graphicx}
\theoremstyle{plain}    
\theoremstyle{plain}    
\theoremstyle{plain}    
\theoremstyle{plain}    
\theoremstyle{plain}    
\theoremstyle{plain}    
\theoremstyle{plain}    
\theoremstyle{plain}    
\theoremstyle{plain}    
\theoremstyle{plain}    
\theoremstyle{plain}    
\theoremstyle{plain}

\input{cmechabbrev}

\newcommand{\kB}{k_\text{B}}  % Boltzmann's constant

\newcommand {\eps} {\epsilon}

\newcommand {\emin} {\epsilon^2_{J\text{min}}}
\newcommand {\TGGL} {\epsilon^2_{\text{TGGL}}}
\newcommand {\BS} {\epsilon^2_{\text{BS}}}
\newcommand {\HVV} {\epsilon^2_{\text{HVV}}}

\newcommand {\egauss} {\epsilon^2_{\text{Gaussian}}}
\newcommand {\edisc} {\epsilon^2_{\text{Discrete}}}

\def \US {U^{\text{store}}}
\def \UC {U^{\text{comp}}}

\begin{document}

\title{The Thermodynamic Uncertainty Theorem}

\author{Kyle J. Ray}
\email{kylejray@gmail.com }
\affiliation{Complexity Sciences Center and Physics Department,
University of California at Davis, One Shields Avenue, Davis, CA 95616, USA}

\author{Alexander B. Boyd}
\email{alecboy@gmail.com}
\affiliation{Corresponding author}
\affiliation{The Division of Physics, Mathematics and Astronomy, California Institute of Technology, Pasadena, CA 91125, USA}
\affiliation{School of Physics, Trinity College Dublin, College Green, Dublin 2, D02 PN40, Ireland}

\author{Giacomo Guarnieri}
\email{giacomo.guarnieri@fu-berlin.de}
\affiliation{Dahlem Center for Complex Quantum Systems,
Freie Universit\"{a}t Berlin, 14195 Berlin, Germany}

\author{James P. Crutchfield}
\email{chaos@ucdavis.edu}
\affiliation{Complexity Sciences Center and Physics Department,
University of California at Davis, One Shields Avenue, Davis, CA 95616}

%\date{\today}

\bibliographystyle{hunsrt_custom}

% ************************* ABSTRACT *************************
\begin{abstract}
Thermodynamic uncertainty relations (TURs) 
express a fundamental tradeoff between the precision (inverse scaled variance) of any thermodynamic current by functionals of the average entropy production. Relying on purely variational arguments, we significantly extend these inequalities by incorporating and analyzing the impact of higher statistical cumulants of entropy production within a general framework of time-symmetrically controlled computation. This allows us to derive an exact expression for the current that achieves the minimum scaled variance, for which the TUR bound tightens to an equality that we name \emph{Thermodynamic Uncertainty Theorem} (TUT). Importantly, both the minimum scaled variance current and the TUT are functionals of the stochastic entropy production, thus retaining the impact of its higher moments. In particular, our results show that, beyond the average, the entropy production distribution's higher moments have a significant effect on any current's precision. This is made explicit via a thorough numerical analysis of swap and reset computations that quantitatively compares the TUT against previous generalized TURs. Our results demonstrate how to interpolate between previously-established bounds and how to identify the most relevant TUR bounds in different nonequilibrium regimes.
\end{abstract}



%\preprint{arxiv.org:XXXX.XXXXX [cond-mat.stat-mech]}

\maketitle

% ****************************************************************

\setstretch{1.1}
\paragraph*{Introduction.}

The past few decades witnessed substantial technological advances in miniaturization that, today, have culminated in experimental realizations of nanoscale thermal machines ~\cite{dubi2011colloquium,blickle2012realization,martinez2016brownian,rossnagel2016single,josefsson2018quantum,maslennikov2019quantum,peterson2019experimental,von_Lindenfels_2019}.  These devices exhibit three fundamental features. First, they operate under nonequilibrium conditions, i.e., either by keeping the system in a nonequilibrium steady-state by means of voltage or temperature biases or through the application of an external time-dependent control protocol. This implies that a certain amount of entropy production $\Sigma$---which quantifies the amount of irreversible dissipation associated with nonequilibrium processes---is always generated~\cite{LandiReview}.  Crucially, the entropy production limits heat engine and refrigerator performance, constrains the physical mechanisms underlying complex biological functioning ~\cite{gnesotto2018broken}, and is the central quantity in Landauer's information erasure---the keystone of the bridge between thermodynamics and communication theory. Second, due to their microscopic nature, the fluctuations of all thermodynamic quantities (such as heat, work, and so on) become as significant as their average values. Last, but not least, the laws of quantum mechanics have important repercussions for fluctuations, which generally have both thermal and quantal origins ~\cite{Feldmann,Plastina,Francica2017,Dann2019}.

Since Onsager's and Kubo's pioneering discovery of the fluctuation-dissipation theorem (FDT) ~\cite{kubo1986brownian,onsager1931reciprocal,kubo1966fluctuation}, determining the universal properties of fluctuations in out-of-equilibrium processes, as well as their role in dissipation, has been a cornerstone of stochastic thermodynamics. In the ‘90s, Jarzynski and Crooks generalized the FDT through the \emph{fluctuation relations} (FRs)~\cite{Gallavotti1995b,Jarz97a,Crooks1998,Tasa00a,Kurc00a,Jarz04a, Andrieux2009,Saito2008,Espo09a,Camp11a,Jarz11a,Hanggi2015}. At the microscopic scale, the FRs refine the famous Second Law of Thermodynamics $\langle\Sigma\rangle \geq 0$ by determining the full distribution of stochastic thermodynamic quantities and thus their fluctuations. That is, the FRs replaced the familiar Second law inequality with an equality from which the Second Law is easily derived through Jensen's inequality.

More recently, a third milestone was crossed by connecting thermodynamic fluctuations out of equilibrium to dissipation. These broad results, called \emph{thermodynamic uncertainty relations} (TURs), were originally discovered in nonequilibrium steady-states of classical time-homogeoneous Markov jump-processes satisfying local detailed balance~\cite{Barato2015,gingrich2016dissipation}. Today, though, TURs have been generalized to finite-time processes~\cite{Pietzonka2017a,Dechant2018,horowitz2017proof}, periodically-driven systems ~\cite{BaratoNJP2018,Holubec2018PRL,proesmans2017discrete,VanVu2020,Guarnieri2021PRL,Guarnieri2021PRE,Falasco2020}, Markovian quantum systems undergoing Lindblad dynamics~\cite{hasegawa2020quantum,hasegawa2021thermodynamic,VuSasa2022}, and autonomous classical~\cite{Dechant2018,dechant2020fluctuation} and quantum ~\cite{MacIeszczak2018,dechant2020fluctuation,Guar19c,SegalAgarwalla,GerryAgarwalla,LiuSegal2021PRE} systems in steady-states close to linear response.

In all these, TURs bound the fluctuations of any (time-reversal anti-symmetric stochastic) thermodynamic quantity $J$ as a function of the \textit{average} entropy production $\langle \Sigma \rangle$:
\begin{align}
\label{BS}
\eps^2_J \equiv \frac{\text{var}(J)}{\langle J \rangle^2}  \geq f(\langle \Sigma \rangle)
 ~.
\end{align} 
with $\langle J\rangle$ and $\text{var}(J) = \langle J^2 \rangle - \langle J \rangle^2$ being the average and variance of $J$, respectively. In this way, the scaled variance $\eps^2_J$ can be seen as the inverse of current $J$'s signal-to-noise ratio or precision. Since $f$ is generally a monotonically-decreasing function, TURs express the trade-off that increased precision in $J$ inevitably comes at the cost of more dissipation. Such a no-free-lunch statement echoes that from the above-mentioned Second Law. It differs critically, however, as it includes fluctuations of many thermodynamic quantities of interest.

Our work significantly advances this line of inquiry by analyzing the impact of higher statistical moments of the entropy production on the signal-to-noise ratio of thermodynamic currents $J$ in two state-of-the-art scenarios; namely, bit swap and reset protocols. We quantify the impact of such higher moments onto the r.h.s. of Eq.~\eqref{BS}---i.e. $f(\langle \Sigma \rangle)$---leading to replacing the latter with:
\begin{align}
\label{eq:TUT}
    \eps^2_J \geq g(\langle f_{\mathrm{min}}(\Sigma) \rangle),
\end{align}
where $g(x) \equiv x^{-1}-1$. With this, the bound becomes a functional of the \textit{stochastic} entropy production $\Sigma$ distribution. And so, critically, it accounts for all higher moments.

Notably, these higher moments have become a focus of  attention \cite{Salazar2022} since they are particularly germane to dissipation management in nanoscale devices---devices that must be designed to tolerate large and potentially destructive fluctuations. For example, entropy production variance and skewness determine the probability of experiencing a trajectory (i.e., an experimental run) that generates extreme dissipated heat flowing through the system. This, naturally, can damage or disrupt the operation of these new classes of microscopic quantum hardware~\cite{Miller2020Landauer}. 

In both examples, we compute Eq.~\eqref{eq:TUT}, and quantitatively compare it against several different previously derived TUR bounds.
In particular, it is important to stress that $g(\langle f_{\mathrm{min}}(\Sigma) \rangle )$ appearing in our Eq.~\eqref{eq:TUT} agrees and coincides with a recent result obtained in Ref. \cite{falasco2020unifying}, albeit via a completely different derivation. Our approach, moreover, focuses on realizing the bound by also finding an explicit expression for the \emph{minimum-variance current} $J_{\mathrm{min}}(\vec{s})$ that saturates Eq.~\eqref{eq:TUT}.

This minimum depends sensitively on the entropy production's higher-order fluctuations. In much the same way that fluctuation theorems \cite{Jarz97a,Jarz00a,Croo99a} reframe the Second Law from an inequality to an equality, the TUT replaces the bounds set by TUR with a saturable equality. Applying the TUT to thermodynamic simulations of fundamental bit swap and reset computations, we demonstrate that current fluctuations can depart substantially from previous bounds set by TURs.

\paragraph*{Minimum Scaled Variance Current.}

Currents $J$ are observations of a system $\mathcal{S}$ that flip sign under time reversal: When a movie of a flowing river is reversed, the positive currents become negative and visa versa.  Formally, a system trajectory is the sequence $\vec{s} \equiv s_0s_{dt} \cdots s_{\tau-dt}s_\tau$, where each $s_{t} \in \mathcal{S}$ is the system's state at time $t$.  The current associated with a reversed trajectory $R(\vec{s})\equiv s^\dagger_\tau s_{\tau-dt}^\dagger \cdots s_{dt}^\dagger s_0^\dagger $ is minus the current of the forward trajectory:
\begin{align}
J(R(\vec{s}))=-J(\vec{s}).
\end{align}
The system $\mathcal{S}$ may be influenced by an external control parameter $\lambda_t$ at every time $t$, thereby performing a computation over the time interval $t \in (0,\tau)$. Under time-symmetric control $\lambda^\dagger_{\tau-t}=\lambda_t$ and conjugation of the distribution under the operation $\Pr(s,\tau)=\Pr(s^\dagger,0)$ the probability of a reverse trajectory is exponentially damped by the entropy production \cite{Hase19a}:
\begin{align}
\Pr(R(\vec{s}),-\Sigma)=e^{-\Sigma}\Pr(\vec{s},\Sigma).
\label{eq:NESSFT}
\end{align}
This is the Detailed Fluctuation Theorem (DFT) for a Time-Symmetrically Controlled Computation (TSCC). (Reference \cite{Supp22a} \ref{app:TSCC DFT} details its derivation and scope.) This DFT ~\cite{Evan02a, Croo99a, Jarz00a} includes NESS dynamics for which the control parameter is constant $\lambda_t=\lambda_{t'}$.  It can also describe, as explored here, computations that begin in equilibrium and are then allowed to relax after the application of a time-symmetric control signal. These latter symmetries are ubiquitous in computing \cite{Riec19b}.

The symmetry imposed by the TSCC imbues $J$'s statistics with special properties in stochastic nonequilibrium systems when compared with the entropy production $\Sigma$. To address this, we derive the exact form of the current $J_\text{min}$ that achieves minimum scaled variance for any TSCC.

Given a TSCC operating over the time interval $[0,\tau]$, described by probability distribution $\Pr(\vec{s},\Sigma)$ over state trajectories $\vec{s}$ and entropy productions $\Sigma$, our task is to find a current function $J(\vec{s})=-J(R(\vec{s}))$ of the state trajectories $\vec{s}$ that minimizes this scaled variance. The proof that this can be done is found in Ref. \cite{Supp22a}. We provide a short overview here.

The demonstration develops over two steps. First, we identify a special class of \emph{entropy-conditioned currents}, that can be expressed as functions of the entropy production, containing a current with the minimum scaled variance.  Then, we use the TSCC detailed fluctuation theorem Eq. (\ref{eq:NESSFT}) and the calculus of variations to derive the primary result---the exact form of this minimum current:
\begin{align}
\label{eq:JMin}
\boxed{J_\text{min}(\vec{s})  = \frac{\langle J^2_\text{min} \rangle}{\langle J_\text{min} \rangle} \tanh\left(\Sigma(\vec{s})/2\right)} .
\end{align}
While Eq. (\ref{eq:JMin}) may appear self-referential, it is worth noting that any real constant may be chosen for the ratio $\frac{\langle J^2_\text{min} \rangle}{\langle J_\text{min}\rangle}$. The resulting current will achieve the minimum variance as dictated by the following \emph{Thermodynamic Uncertainty Theorem}.

\begin{theorem}[Thermodynamic Uncertainty Theorem]
In a TSCC, the scaled variance of any current $J$ is bounded below by the the scaled variance of $J_\text{min}$, given by:
\begin{align}
\label{eq:EMin}
\boxed{ \emin= \frac{1}{\langle \tanh(\Sigma/2) \rangle}-1}.
\end{align}
\end{theorem}
Since this minimum is achievable by a well-defined current $J_\text{min}$, it sets the tightest possible bound that can be determined using all moments of the entropy production distribution. We note that this expression contrasts with other past TURs \cite{Bara15a,Horo17a,Hase19a,Timp19a} in that it is an \emph{equality} rather than an inequality.  It specifies the achievable minimum scaled variance since it depends on the underlying physical system's entropy production distribution $\Pr(\Sigma)$.  

Equation~\eqref{eq:EMin} agrees and coincides with Eq.~(16) of Ref.~\cite{falasco2020unifying}, which was derived through a completely different method; namely, by means of a ``Hilbert uncertainty relation'' that leverages the reproducing element implied by the Riesz representation theorem for any Hilbert space equipped with an appropriate inner product. This uncertainty relation was shown to reduce to the an equation of the form of Eq.~\eqref{eq:TUT}. Our approach is complementary as it investigates how the relation stems from thermodynamics through a detailed fluctuation theorem (DFT). While reaching the same conclusion, our method has two additional merits. First, it clearly highlights its relation to other previously generalized TUR bounds, also obtained from the requirement of a DFT.
Thus, it gives a quantitative comparisons, as shown shortly. Second, it determines a physically realizable \emph{minimum-variance currents} $J_{\mathrm{min}}(\vec{s})$ that saturates Eq.~\eqref{eq:EMin}.

\paragraph*{Comparison to Past Uncertainty Relations.}

With a blossoming variety uncertainty relations \cite{Barato2015,gingrich2016dissipation,Pietzonka2017a,Dechant2018,horowitz2017proof,BaratoNJP2018,Holubec2018PRL,proesmans2017discrete,VanVu2020,Guarnieri2021PRL,Guarnieri2021PRE,Falasco2020,hasegawa2020quantum,hasegawa2021thermodynamic,VuSasa2022,Dechant2018,dechant2020fluctuation,MacIeszczak2018,dechant2020fluctuation,Guar19c}, it is natural to ask for direct comparisons. Reference \cite{Supp22a}  \ref{app:Background} provides a thorough summary of previous TUR results, but here we summarize the relevant comparisons. Barato and Seifert \cite{Bara15a} derived the first TUR, finding that precision could not be maximized without a corresponding increase in the average entropy production:
\begin{align}
\eps^2_J \geq \frac{2}{\langle \Sigma \rangle} \equiv \BS
  ~,
\end{align}
where the subscripts on the latter refer to the original authors---the Barato-Seifert (BS) bound on scaled variance.

\begin{figure}
\includegraphics[width=\columnwidth]{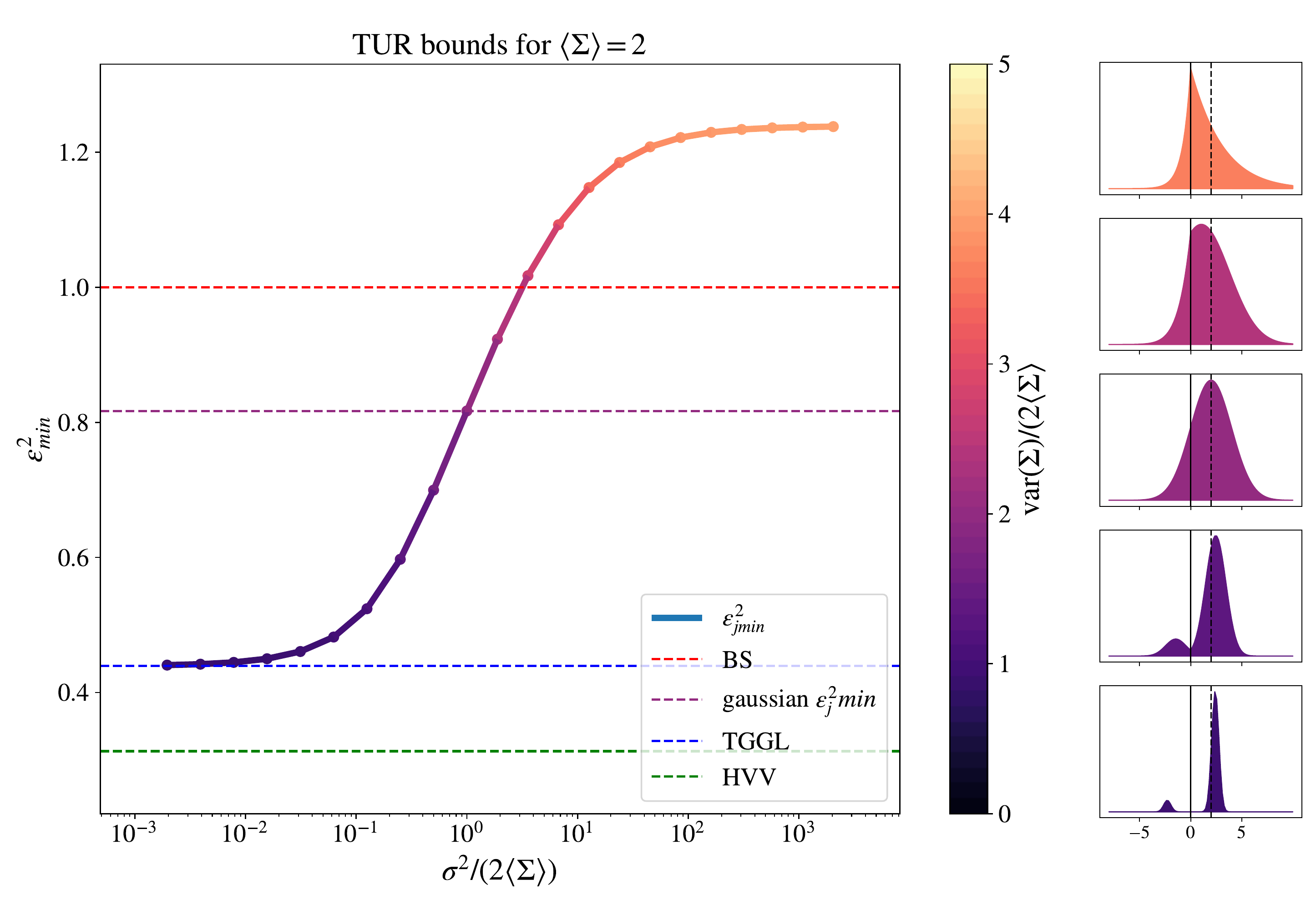}
\caption{Thermodynamic Uncertainty Theorem and TURs: Three dashed lines show previous TURS---$\BS$, $\HVV$, and $\TGGL$---all functions of average entropy production $\langle \Sigma \rangle$. While they make nearly identical predictions for small entropy production, they diverge as entropy increases, setting very different bounds for average entropy production as low as $\langle \Sigma \rangle = 2k_\text{B}$. In contrast, the minimum scaled variance $\emin$ is not strictly a function of average entropy.  The entropy distribution $\Pr(\Sigma|\mu,\sigma^2)$ depends on the variance parameter $\sigma^2$ and is displayed on a sliding scale from high to low variance (from dark to light). $\sigma^2$ ranges from $\approx 8\times 10^{-3}$ to $8\times 10^3$. While $\mu$ is adjusted to keep $\langle \Sigma \rangle$ fixed. This yields entropy distributions of the form shown on the right.  The lowest values of $\Sigma^2$ and $\text{var}(\Sigma)$ closely match $\TGGL$.  As $\sigma^2$ increases, the variance of the entropy production increases and the curve become lighter, achieving and surpassing the dashed line for $\BS$.  Between these two extremes, there is a purple dashed line that gives the minimum scaled variance of normal entropy distributions.}
\label{fig:Jmin} 
\end{figure}

Further exploration found that detailed fluctuation theorems \cite{Jarz00a,Croo99a} can be used to establish modified thermodynamic uncertainty relations. Hasegawa and Van Vu \cite{Hase19a} as well as Timparano, Guarnieri, Goold, and Landi \cite{Timp19a} used Eq. (\ref{eq:NESSFT}) to demonstrate that the scaled variance is bounded below by:
\begin{align}
\eps^2_J \geq \frac{2}{e^{\langle \Sigma \rangle}-1} &\equiv \HVV \\
\eps_J^2 \geq  \text{csch}^2[g(\langle \Sigma \rangle/2)] &\equiv \TGGL,
\end{align}
where $g(x)$ is the inverse of $x\tanh (x)$ and we have again labeled the bounds for the authors. The $\TGGL$ bound is the tightest possible bound on scaled variance that can be determined from the average entropy \cite{Timp19a}.  Comparing these bounds independent of the TUT, one can see in Figs. \ref{fig:Jmin} and \ref{fig:JminSim} that they are ordered:
\begin{align}
\BS(\langle \Sigma \rangle)> \TGGL(\langle \Sigma \rangle)>  \HVV (\langle \Sigma \rangle).
\end{align}
(See also Ref. \cite{Supp22a} \ref{app:BoundOrdering} for a proof.)
Note that, since $\TGGL$ and $\HVV$ were derived from the TSCC DFT, which is our starting point as well, the minimum scaled variance is bounded below by these TURS, but not necessarily $\BS$.

With $\emin$'s exact form determined, though, a natural next question is how close the previous bounds, all depending only on the average entropy production $\langle \Sigma \rangle$, are to the actual minimum.

Fortunately, Timpanaro et al. \cite{Timp19a} also showed that a particular bimodal distribution:
\begin{align*}
\Pr_\text{min}(\Sigma) \propto \delta( \Sigma-a) + e^{-a} \delta ( \Sigma+ a)
\end{align*}
achieves their lower bound $\TGGL$. This is the simplest distribution satisfying $\Pr(-\Sigma)=e^{-\Sigma} \Pr(\Sigma)$. That is, it consists of a delta function at entropy production $\Sigma=a$ and then contains a mirror of that entropy production at $\Sigma=-a$ reduced by the exponential factor $e^{-a}$.  Any other NESS entropy distribution can be constructed from a superposition of such distributions.

We investigate our new minimum scaled variance by exploring a variety of possible distributions.  We take a similar strategy, breaking the entropy distribution into the piecewise function:
\begin{align}
\Pr(\Sigma| \mu, \sigma^2)=n(\mu,\sigma^2) \begin{cases}
e^{\frac{(\Sigma-\mu)^2}{2 \sigma^2}} & \text{ if } \Sigma \geq 0 \\
e^{\Sigma}e^{\frac{(-\Sigma-\mu)^2}{2 \sigma^2}} & \text{ if } \Sigma < 0 
\end{cases}~.
\end{align}
Here, $n( \mu,\sigma^2)= \int_0^\infty e^{\frac{(\Sigma-\mu)^2}{2 \sigma^2}} d\Sigma + \int_{-\infty}^0 e^{\Sigma}e^{\frac{(-\Sigma-\mu)^2}{2 \sigma^2}} d\Sigma$ is the normalization factor.  In essence, our probability distribution is a normal distribution with average $\mu$ and variance $\sigma^2$ over the positive interval. And, the TSCC DFT defines the distribution to be $\Pr(\Sigma)=e^{\Sigma} \Pr(-\Sigma)$ on the negative interval.  

The variance $\sigma^2$ and average $\mu$ of the positive entropy portion of the distribution $\Pr(\Sigma|\mu,\sigma^2)$ define it.  In the limit $\sigma^2 \ll k_B \mu$, the positive entropy distribution is nearly a delta function, and we recover roughly the distribution $\Pr_\text{min}(\Sigma)$ proposed by Timpanaro et al \cite{Timp19a}. We show this on the left side of Fig. \ref{fig:Jmin} where $\sigma^2 \approx 10^{-3} k_B \mu$, corresponding to a two-peaked distribution. In this case, $\emin$ closely matches the bound $\TGGL$, as expected.  However, Fig. \ref{fig:Jmin} also shows that the average entropy production is not the sole determinant of the minimum scaled variance of the current.

As the variance of the positive normal distribution $\sigma^2$ increases, Fig. \ref{fig:Jmin} shows that the minimum scaled variance increases.  Amidst that progression is a special distribution, where $\sigma^2= 2k_B \mu$, for which $\Pr(\Sigma| \mu, \sigma^2)$ is a normal distribution over the full range of entropy production.  It can be quickly shown that the variance for a normal distribution that satisfies the TSCC DFT must be $\sigma^2=2 k_B \mu$. We highlight this special case with a purple dashed line in Fig. \ref{fig:Jmin} and a blue dashed line in Fig. \ref{fig:JminSim}.  NESSs, the most frequently studied subclass of TSCC processes, approach a normal entropy production distribution in the long time limit.  Interestingly, the minimal variance currents of the typical asymptotic behavior in NESSs clearly violate the original TUR given by $\BS$ in the long-time limit.

If we continue beyond the normal distribution to higher variances labeled in lighter colors in Fig. \ref{fig:Jmin}, the minimum scaled variance $\emin$ continues to increase, until it surpasses the bound $\BS$ \cite{Bara15a}.  Thus, by changing the parameter $\sigma^2$ of the NESS distribution $\Pr(\Sigma | \mu, \sigma^2)$ and its variance $\text{var}(\Sigma)$ as well as other higher moments of the entropy distribution $\Pr(\Sigma)$, the TUT interpolates between the TUR set by Timpanaro et al. and that set by Barato and Seifert.  Moreover, we can find entropy distributions that far exceed even Barato and Seifert's bound.

\paragraph*{Thermodynamic Simulations.}

\def \resetN {1366 }
\def \swapN {1193 }

\begin{figure*}
\includegraphics[width=2\columnwidth]{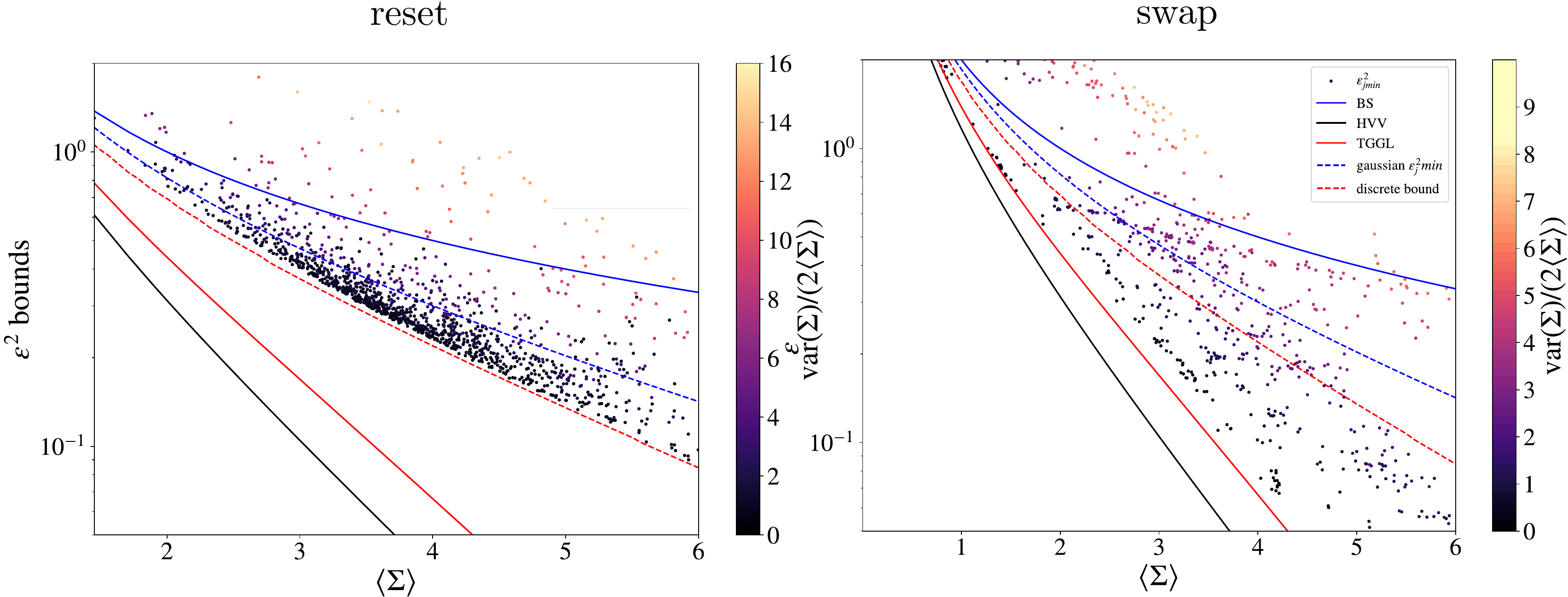}
\caption{Bounds $\BS> \TGGL > \HVV$ in solid lines (blue, red, and black respectively) and specific thermal processes with dashed lines:  The blue dashed line is the minimum scaled variance $\egauss$ of any process that generates a Gaussian entropy production distribution, which is achieved in the long-time limit of NESS processes.  The red dashed line is the minimum scaled variance $\edisc$ of an ideal discrete erasure.  We compare two computational classes to the these bounds. (Left) we plot \resetN different time-symmetric erasures.  As expected (see Ref. \cite{Supp22a} \ref{app:DiscreteBound}) they are bounded by the scaled variance of the ideal discrete erasure $\edisc$, which lies well above the bounds $\TGGL$ and $\HVV$.  A number of erasure operations are well above the minimum $\TGGL$ set by Barato and Seifert. (Right) we plot the result \swapN different bit-flips.  As with the erasure protocol, many computations are above the Barato-Seifert bound.  However, many computations achieve a minimum scaled variance well below the discrete erasure bound $\edisc$.  Many computations are quite close to the strongest possible TUR $\TGGL$, indicating that this theoretical bound is indeed achievable with TSCCs.}
\label{fig:JminSim} 
\end{figure*}

The entropy production distribution $\Pr(\Sigma|\mu, \sigma^2)$ is convenient to examine, but it is not obvious how it can be physically generated.  We now describe two computational protocols that are able to show similar breadth of behavior, but are firmly rooted in dynamical models of physical processes.  Both are TSCCs in that they are implemented with time symmetric control of a potential energy landscape, where the thermal influence of a bath is applied through Langevin dynamics.

First, consider a simple reset protocol.  A system consisting of a single positional variable $x$ in asymmetric double well potential $\US$ is initially set up in equilibrium with a thermal environment at temperature $T$.  If the two metastable states are $A$ and $B$, then take $A$ as the one with a deeper well and higher initial probability.  Then, the energy landscape is tilted and the energy barrier is removed. This computational potential $\UC$ is held so that probability mass flows from $A$ to $B$.  This is a ``reset'' in that it re-initializes the system to the $B$ state \cite{Riec19b,Wims21a}. Ref. \cite{Supp22a} \ref{app:Sim} describes this protocol in detail.

The resulting entropy productions in minimum-variance currents are shown in Fig. \ref{fig:JminSim}. Some computations are considerably less precise than specified by $\BS$, the original TUR, but many lie well below this bound.  As expected, all computations are less precise than the bounds $\TGGL$ and $\HVV$, but none of them come very close to the tightest theoretical bound given by $\TGGL$.  Instead, there is another curve that seems to bound all of these time-symmetric erasures, shown in dashed red.  Reference \cite{Supp22a}  \ref{app:DiscreteBound} derives the bound $\edisc$ from a discrete-state time-symmetric erasure, where the system equilibrates to the tilted energy landscape before the metastable information-storing potential is re-initialized at the end of the computation.

Second, consider the same initial metastable states $A$ and $B$, which are stored in local equilibrium.  Then, instantaneously implement a harmonic potential and hold the energy landscape for half a period of the oscillation.  If the coupling to the thermal environment is weak, then this implements a reliable swap between $A$ and $B$, using momentum as memory to carry the distributions into their new states.  This is a slight modification of momentum computing protocols described in Refs. \cite{Ray21a, Ray22a}.  The entropy and scaled variance of these swaps are shown on the right-hand side of Fig. \ref{fig:JminSim}.  They span the same space of possibilities shown for $\Pr(\Sigma|\mu,\sigma^2)$ in Fig. \ref{fig:Jmin}: some lying above $\BS$ and some TSCCs sitting just above the minimum set by $\TGGL$. 

For both the reset and swap, we see that the minimum scaled variance tends to increase as the variance of the entropy-production distribution increases.  Higher moments of entropy production are critically important in predicting the precision of computations.

\paragraph*{Conclusion.}

We introduced two equalities that provide an explicit expression for the most accurate current in any entropy production distribution that satisfies Eq. (\ref{eq:NESSFT}) and for the minimal scaled variance it achieves. Analogous to the transition from the Second Law inequality to the fluctuation relations of Crooks and Jarzynski, we developed a treatment of the entropy production that recognizes it as a stochastic quantity with proper fluctuations. This treatment yields a result---the Thermodynamic Uncertainty Theorem---that is (i) an equality rather than an inequality and (ii) depends on the stochastic entropy production's fluctuations rather than on only its average value. 

The TUT also allows for straightforward derivations of new system or dynamic specific TURs. Our simulations also demonstrate this property: The overdamped reset operations adhere to their TUT-derived bound $\edisc$, while the underdamped swap operations are free of $\edisc$ and can approach the $\epsilon^2_\text{TGGL}$ bound. Tuning the higher moments of entropy production allows us to interpolate between the maximum precision $\epsilon^2_\text{TGGL}$ set by Timpanaro et al. \cite{Timp19a} and the original TUR $\epsilon^2_\text{BS}=\frac{2}{\langle \Sigma \rangle}$ \cite{Bara15a}.  Nonequilibrium steady states generally approach a Gaussian entropy production distribution, so their higher moments produce currents that are more precise than predicted by $\epsilon^2_\text{BS}$ but less than $\epsilon^2_\text{TGGL}$.

We also considered time-symmetrically controlled computation. Our simulations show that there are both reset and swap operations for which the minimum variance current is even less precise than predicted by $\epsilon^2_\text{BS}$. While the relationship is not exact, these results indicate that, independent of the average entropy production, higher variance in the entropy production leads to less precise currents. This analysis opens up the opportunity to investigate, in more detail, the specific effects that these higher moments impose upon the entropy production. Equation (\ref{appeq:EminMoments}) gives the formal relationship, but it remains to be seen how these effects manifest in practice.

\paragraph*{Acknowledgements. --} We thank John Goold and Felix Binder for helpful discussions and Gianmaria Falasco, Massimiliano Esposito, and Jean-Charles Delvenne for alerting us to Ref.~\cite{falasco2020unifying}. The authors thank the Telluride Science Research Center for hospitality during visits and the participants of the Information Engines Workshops there. J. P. C. acknowledges the kind hospitality of the Santa Fe Institute, Institute for Advanced Study at the University of Amsterdam, and California Institute of Technology.  A. B. B. acknowledges support from  the Templeton World Charity Foundation Power of Information fellowship TWCF0337 and TWCF0560 and from the Foundational Questions Institute and Fetzer Franklin Fund, a donor advised fund of the Silicon Valley Community Foundation grant number FQXi-RFP-IPW-1910.  J. P. C. and K. J. R. acknowledge support by, or in part by, FQXi Grant number FQXi-RFP-IPW-1902 and U.S. Army Research Laboratory and the U.S. Army Research Office under grants W911NF-21-1-0048 and W911NF-18-1-0028.  G. G. acknowledges support from FQXi and DFG FOR2724 and also from the European Union Horizon 2020 research and innovation programme under the Marie Sklodowska-Curie grant agreement No. 101026667.

%=================================================================
% References:
%=================================================================

\bibliography{chaos,library2}

\pagebreak
\widetext

\newpage 
\begin{center}
\vskip0.5cm
{\Large Supplemental Material for: \\[0.2in]
The Thermodynamic Uncertainty Theorem}\\[0.2in]
{Kyle J. Ray, Alexander B. Boyd, Giacomo Guarnieri, and James P. Crutchfield}
\end{center}

%%%%%%%%%% Prefix a "S" to all equations, figures, tables and reset the counter %%%%%%%%%%
\setcounter{section}{0}
\setcounter{equation}{0}
\setcounter{figure}{0}
\setcounter{table}{0}
\setcounter{page}{1}
\renewcommand{\theequation}{S\arabic{equation}}
\renewcommand{\thefigure}{S\arabic{figure}}

\section{Time Symmetrically Controlled Computation Detailed Fluctuation Theorem}
\label{app:TSCC DFT}

While Crooks \cite{Croo99a} originally derived the Detailed Fluctuation Theorem (DFT) within the context of detailed balanced Markov dynamics, Jarzynski \cite{Jarz00a} provides an elegant proof of a generalization of this result with a rather appealing set-up.  With a system $\mathcal{S}$ that is controlled by an external parameter and weakly coupled to a collection of thermal reservoirs (each specified by its respective temperature $T$), he proved that the joint probability of a state trajectory $\vec{s}\equiv s_0 s_{t_1} \cdots s_{t_{N-2}} s_\tau$, and the environmental entropy production, defined as the heat entropy generated in each reservoir:
\begin{align}
    Q \equiv \sum_{T}\frac{Q_T}{T},
\end{align}
is exponentially damped under time reversal.  This framework can be used to do a wide variety of computations over the time interval $t \in (0,\tau)$ and can operate without the constraints of detailed balance, because there are multiple reservoirs.

Specifically, using random variable notation where capital letters denote the random variable and lower case letters denote their realizations, Ref. \cite{Jarz00a} shows that:
\begin{align}
\label{eq:JarzDFT}
\frac{ \Pr(Q=-q,\vec{S}=R(\vec{s})|S_0=s_\tau^\dagger, \vec{\Lambda}=R(\vec{\lambda}))}{\Pr(Q=q,\vec{S}=\vec{s}|S_0=s_0, \vec{\Lambda}=\vec{\lambda})} =e^{-q},
\end{align}
where we set Boltzmann's constant to $k_B=1$ for ease of notation.  Here, $\vec{\lambda}\equiv \lambda_0 \cdots \lambda_\tau$ is the control parameter trajectory over the computation interval $(0,\tau)$.  The time reversal operator $R$ operates on trajectories by reversing the sequence and flipping the sign of time-antisymmetric variables like magnetic fields and momentum $R(\vec{\lambda})\equiv \lambda_\tau^\dagger \cdots \lambda_0^\dagger $.  Beyond weak coupling, Ref. \cite{Jarz00a} assumes Hamiltonian dynamics and that the thermal baths are initially independently distributed in their respective Boltzmann distributions.

This differs from the Crooks fluctuation theorem in that $\vec{s}$ represents a potentially discrete sequence of states taken at potentially irregular times $t_i \in (0, \tau)$ and in that it also included dependence on the entropy production $Q$ in the probability.  However, in the limit where $\vec{s}$ constitutes a nearly complete description of the state trajectory, the environmental entropy production is a deterministic function of the state trajectory, and there is only one heat bath, the results of Refs. \cite{Croo99a,Jarz00a} are similar.

In essence, both of these detailed fluctuation theorems recover thermodynamic properties of a forward experiment by comparing trajectory probabilities to those of a reverse experiment.  The forward experiment is determined by the control protocol $\vec{\lambda}$, which determines the probability of state trajectories and heats conditioned on the initial state $S_0$, but also depends sensitively on the initial distribution of the system $f_0(s_0)$.  $f$ is used to indicate that it is the probability of the forward experiment, and the subscript $0$ indicates initial time $t=0$.  This then provides the instantaneous distribution of the system in the forward experiment at all times $t$:
\begin{align}
    f_t(s)\equiv \sum_{s_0}\Pr(S_t=s|S_0=s_0,\vec{\Lambda}=\vec{\lambda})f_0(s_0).
\end{align}
Similarly, the reverse experiment must be prepared in some distribution $r_0(s_0)$, yielding instantaneous distribution of the system in the reverse experiment at all times:
\begin{align}
    r_t(s)\equiv \Pr(S_t=s|S_0=s_0,\vec{\Lambda}=R(\vec{\lambda}))r_0(s_0).
\end{align}
The state of preparation of the forward and reverse experiment is essential for determining the total entropy production of the system.

The total entropy production due to control $\vec{\lambda}$ in the forward experiment must also include the change in surprisal of the system \cite{Seif05a}:
\begin{align}
    \Sigma=q+ \ln \frac{f_0(s_0)}{f_\tau(s_\tau)}.
\end{align}
Plugging into the DFT in Eq. (\ref{eq:JarzDFT}), we obtain a relation for total entropy production as well:
\begin{align}
\frac{ \Pr(Q=-q,\vec{S}=R(\vec{s})|S_0=s_\tau^\dagger, \vec{\Lambda}=R(\vec{\lambda}))}{\Pr(Q=q,\vec{S}=\vec{s}|S_0=s_0, \vec{\Lambda}=\vec{\lambda})} & =e^{-\Sigma+\ln \frac{f_0(s_0)}{f_\tau(s_\tau)}} \nonumber
\\ \frac{ \Pr(Q=-q,\vec{S}=R(\vec{s})|S_0=s_\tau^\dagger, \vec{\Lambda}=R(\vec{\lambda}))}{\Pr(Q=q,\vec{S}=\vec{s}|S_0=s_0, \vec{\Lambda}=\vec{\lambda})} &\frac{f_\tau(s_\tau)}{f_0(s_0)}  =e^{-\Sigma}.
\end{align}

To parallel Crooks original examination of fluctuation theorems \cite{Croo99a}, we would like to express probabilities of \emph{total} entropy production, rather than \emph{environmental} entropy production.  En route, note that the conditional probability of total entropy production under forward control can be expressed:
\begin{align}
    \Pr(\Delta S_\text{tot}= \Sigma| Q=q, \vec{S}=\vec{s} , \vec{\Lambda}=\vec{\lambda})= \delta_{\Sigma ,q+ \ln \frac{f_0(s_0)}{f_\tau(s_\tau)}}.
\end{align}
We can then evaluate:
\begin{align}
\label{eq:TSCCderiv}
     \Pr(\Delta S_\text{tot}=\Sigma,\vec{S}=\vec{s}|S_0=s_0, \vec{\Lambda}=\vec{\lambda}) f_0(s_0) & =
 \sum_{q} \delta_{\Sigma ,q+ \ln \frac{f_0(s_0)}{f_\tau(s_\tau,)}}\Pr(Q=q,\vec{S}=\vec{s}|S_0=s_0, \vec{\Lambda}=\vec{\lambda})f_0(s_0)
\\      \Pr(\Delta S_\text{tot}=\Sigma,\vec{S}=\vec{s}| \vec{\Lambda}=\vec{\lambda}) & =
 e^\Sigma \sum_{q} \delta_{-\Sigma ,-q+ \ln \frac{f_\tau(s_\tau)}{f_0(s_0)}}\Pr(Q=-q,\vec{S}=R(\vec{s})|S_0=s_\tau^\dagger, \vec{\Lambda}=R(\vec{\lambda}))f_\tau(s_\tau), \nonumber
 \\      \Pr(\Delta S_\text{tot}=\Sigma,\vec{S}=\vec{s}| \vec{\Lambda}=\vec{\lambda}) & =
 e^\Sigma \sum_{q} \delta_{-\Sigma ,-q+ \ln \frac{f_\tau(s_\tau)}{f_0(s_0)}}\Pr(Q=-q,\vec{S}=R(\vec{s})| \vec{\Lambda}=R(\vec{\lambda})), \nonumber
\end{align}
where we introduced the unconditioned process that comes from forward control $\vec{\lambda}$ and starting in the initial distribution $f_0(s_0)$:
\begin{align}
& \Pr(\Delta S_\text{tot}=\Sigma,\vec{S}=\vec{s}| \vec{\Lambda}=\vec{\lambda})
\\ & \equiv  \Pr(\Delta S_\text{tot}=\Sigma,\vec{S}=\vec{s}|S_0=s_0, \vec{\Lambda}=\vec{\lambda}) f_0(s_0). \nonumber
\end{align}
We also introduced unconditioned process of the reverse experiment that comes from initializing the system in the conjugate of the final distribution of the forward experiment $r_0(s) = f_\tau(s^\dagger)$ and reversing the control sequence:
\begin{align}
   & \Pr(Q=q,\vec{S}=\vec{s}| \vec{\Lambda}=R(\vec{\lambda})) \\ & \equiv \Pr(Q=q,\vec{S}=\vec{s}|S_0=s_0, \vec{\Lambda}=R(\vec{\lambda}))r_0(s_0). \nonumber
\end{align}
Plugging in $-q$ for the environmental entropy production and $R(\vec{s})$ for the state sequence yields the right-hand term in the last line of Eq. (\ref{eq:TSCCderiv}).

We want to evaluate the probability of entropy production in the reverse experiment as well.  Denoting the final system distribution that results from the reverse experiment as $r_\tau(s)$, the total entropy production is:
\begin{align}
    \Sigma= q + \ln \frac{r_0(s_0)}{r_\tau(s_\tau)},
\end{align}
meaning that:
\begin{align}
   & \Pr(\Delta S_\text{tot}=\Sigma,\vec{S}=\vec{s}| \vec{\Lambda}=R(\vec{\lambda}))
    \\ & =\sum_{q} \delta_{ \Sigma, q + \ln \frac{r_0(s_0)}{r_\tau(s_\tau)}} \Pr(Q=q,\vec{S}=\vec{s}| \vec{\Lambda}=R(\vec{\lambda})). \nonumber
\end{align}
Comparing with Eq. (\ref{eq:TSCCderiv}), we evaluate the probability of realizing the reverse sequence $R(\vec{s})$ and minus the entropy production $-\Sigma$, while summing over $-q$ rather than $q$:
\begin{align}
   & \Pr(\Delta S_\text{tot}=-\Sigma,\vec{S}=R(\vec{s})| \vec{\Lambda}=R(\vec{\lambda}))
    \\ & =\sum_{q} \delta_{ -\Sigma, -q + \ln \frac{r_0(s_\tau^\dagger)}{r_\tau(s_0^\dagger)}} \Pr(Q=-q,\vec{S}=R(\vec{s})| \vec{\Lambda}=R(\vec{\lambda})). \nonumber
\end{align}
If we want to satisfy a parallel equality to the Crooks fluctuation theorem:
\begin{align}
 \label{Eq:Extension}
   & \Pr(\Delta S_\text{tot}=\Sigma,\vec{S}=\vec{s}| \vec{\Lambda}=\vec{\lambda}) 
    \\ & =
 e^\Sigma \Pr(\Delta S_\text{tot}=-\Sigma,\vec{S}=R(\vec{s})| \vec{\Lambda}=R(\vec{\lambda})), \nonumber
 \end{align}
 for all reverse distributions $\Pr(Q=q,\vec{S}=\vec{s}| \vec{\Lambda}=R(\vec{\lambda}))$, then it must be true that:
 \begin{align}
\delta_{-\Sigma ,-q+ \ln \frac{f_\tau(s_\tau)}{f_0(s_0)}}=\delta_{ -\Sigma, -q + \ln \frac{r_0(s_\tau^\dagger)}{r_\tau(s_0^\dagger)}}.
\end{align}
Thus, it must be true that:
\begin{align}
  \frac{f_\tau(s_\tau)}{f_0(s_0)}= \frac{r_0(s_\tau^\dagger)}{r_\tau(s_0^\dagger)}.
\end{align}
Since we have chosen $r_0(s) = f_\tau(s^\dagger)$, the final system distribution of the system in the reverse experiment must be the same as the conjugate of the initial distribution in the forward experiment:
\begin{align}
\label{Eq:Cond}
    r_\tau(s)=f_0(s^\dagger).
\end{align}
This nuance was recongized in Crooks's original result \cite{Croo99a}, though it has sometimes been lost in translation.

There are a litany of different cases in which $ r_\tau(s)=f_0(s^\dagger)$ is violated, meaning that the extension of Crooks fluctuation theorem shown in Eq. (\ref{Eq:Extension}) doesn't necessarily hold.  For instance, if we choose a control protocol that erases a uniform distribution over $\{\uparrow, \downarrow \}$, taking $f_0(s)=1/2$ to $f_\tau(s)= \delta_{s,\downarrow}$, but does so time symmetrically, then $r_\tau(s)=\delta_{s,\downarrow} \neq f(s^\dagger)=1/2$.  

However, in this special case of time-symmetric control ($R(\vec{\lambda})=\vec{\lambda}$) we see a useful simplification.  If the final distribution of the forward experiment happens to conjugate the initial state distribution $f_\tau(s)=f_0(s^\dagger)$, then the initial distribution of the reverse control experiment is the same for the forward control $r_0(s)=f_0(s)$.  The resulting final system distribution of reverse control is also the same $r_\tau(s)=f_\tau(s)$, because the forward and reverse experiment apply the same control protocol to the same initial distribution.  This guarantees the condition in Eq. (\ref{Eq:Cond}) that $r_\tau(s)=f_0(s^\dagger)$, and so we see that:
\begin{align}
\Pr(\Sigma,\vec{s}) = e^{\Sigma}\Pr(-\Sigma,R(\vec{s})), 
\end{align}
where we have shortened to the notation $\Pr(\Sigma,\vec{s})\equiv \Pr(\Delta S_\text{tot}=\Sigma,\vec{S}=\vec{s}| \vec{\Lambda}=\vec{\lambda})$ for convenience.  This is the Time-Symmetric Control Computation Detailed Fluctuation Theorem (TSCC DFT), which comes from the assumption of time-symmetric control $\vec{\lambda}=R(\vec{\lambda})$, and conjugation of the state distribution under the computation $f_\tau(s) = f_0(s^\dagger)$.  

This differs slightly but importantly from the conditions for the TSCC DFT described by Hasegawa and Van Vu \cite{Hase19a}, who instead assume that the starting state is preserved ($f_\tau(s) = f_0(s)$) under the computation. This assumption may fail to produce the desired fluctuation theorem if the control acts differently on the conjugate of the initial distribution.  In this case, the computation has a different effect on the starting distribution of the reverse protocol $r_0(s)\equiv f_\tau(s^\dagger)$, which is also equal to the conjugate of the initial distribution of the forward experiment, because $f_\tau(s^\dagger)=f_0(s^\dagger)$.  The distribution won't be preserved under the computation $r_\tau(s) \neq r_0(s)$.  Because $r_0(s)=f_0(s^\dagger)$, the condition for the DFT in Eq. \ref{Eq:Cond} isn't met.  The assumptions of Ref. \cite{Hase19a} are sufficient when the initial distribution is time reversal symmetric $f_0(s^\dagger)=f_0(s)$, which is often the case in biochemical systems, where the time-antisymmetric momentum variables are thermalized with respect to a single heat bath.  This is also the case in many Nonequilibrium Steady States.

\section{Minimum Current Derivation}
\label{app:Proof}

\subsection{Entropy-Conditioned Current}

Currents are functions of state trajectories, so they can be applied to the distribution that characterizes the TSCC $\Pr(\vec{s}, \Sigma)$ to derive a three-variable joint distribution over currents, state trajectories, and entropy productions:
\begin{align*}
\Pr(J,\vec{s},\Sigma)\equiv \delta_{J,J(\vec{s})}\Pr(\vec{s},\Sigma).
\end{align*}
From this, one can determine a variety of marginal distributions.  Most relevant to us is the joint probability of currents and entropy productions $\Pr(J,\Sigma)$, and the probability of a current given an entropy production $\Pr(J|\Sigma)$.  From the latter, we define the entropy-conditioned current $j(\Sigma)$ as the average current in the system given that the entropy $\Sigma$ was dissipated in the system:
\begin{align}
j(\Sigma) \equiv \sum_{J} J \Pr(J|\Sigma).
\end{align}
If the entropy production is a function of the state trajectory, as is the case for systems satisfying local detailed balance, then we can use the entropy conditioned current to define a new function of the trajectories:
\begin{align}
J'(\vec{s})\equiv j(\Sigma(\vec{s})).
\end{align}
$J'$ is a well-defined current within the system, because $J'(R(\vec{s}))=-J'(\vec{s})$, as shown in a later section of the Supplementary Material. \ref{app:Well-defined current}

The newly defined entropy-conditioned current has the convenient property that it's average $\langle J' \rangle \equiv \langle j \rangle$ is the same as for the current that was used to define it:
\begin{align*}
\langle J' \rangle &  = \sum_{\Sigma}  \Pr(\Sigma) j(\Sigma)
\\ & =\sum_{\Sigma}  \Pr(\Sigma) \sum_{J} J \Pr(J|\Sigma)
\\ & =  \sum_{\Sigma ,J} J \Pr(J,\Sigma)
\\ & =  \sum_{J} J \Pr(J)
\\ & \equiv \langle J \rangle.
\end{align*}
On the other hand, the variance of the entropy-conditioned current is not the same.  When we evaluate the average square of the newly defined current:
\begin{align*}
\langle J'^2 \rangle & =\sum_\Sigma \Pr(\Sigma) j(\Sigma)^2
\\ & =\sum_\Sigma \Pr(\Sigma)\left( \sum_{J} J \Pr(J|\Sigma) \right)^2,
\end{align*}
we can apply Jensen's inequality $\left( \sum_{J} J \Pr(J|\Sigma) \right)^2 \leq \left( \sum_{J} J^2 \Pr(J|\Sigma) \right)$ to show that: 
\begin{align*}
\\ \langle J'^2 \rangle &  \leq \sum_\Sigma \Pr(\Sigma)\left( \sum_{J} J^2 \Pr(J|\Sigma) \right)
\\ & = \sum_{J,\Sigma} \Pr(J ,\Sigma) J^2
\\ & = \langle J^2 \rangle.
\end{align*}
As a result, the entropy-conditioned current produces scaled variance that is less than or equal to the current that was used to define it:
\begin{align}
\epsilon^2_{J} \geq \epsilon^2_{J'}.
\end{align}
Thus, we need only consider currents which are functions of the entropy production in order to find the minimum-variance current.

\subsection{Well-defined current}
\label{app:Well-defined current}

Let us define function $J'$ of state trajectories in terms of the entropy-conditioned current:
\begin{align*}
    J'(\vec{s})\equiv j(\Sigma(\vec{s})),
\end{align*}
where the entropy conditioned current is defined:
\begin{align*}
    j(\Sigma) \equiv \sum_J J \Pr(J|\Sigma).
\end{align*}
Note that we can evaluate $J'$ for the time-reversal of a trajectory:
\begin{align*}
 J'(R(\vec{s}))=j(\Sigma(R(\vec{s}))).
\end{align*}

The TSCC fluctuation theorem:
\begin{align*}
\Pr(R(\vec{s}),-\Sigma)=e^{-\Sigma}\Pr(\vec{s},\Sigma),
\end{align*}
implies both a marginalized version:
\begin{align*}
    \Pr(-\Sigma)=e^{-\Sigma}\Pr(\Sigma),
\end{align*}
as well as equality of the conditional probabilities:
\begin{align*}
\Pr(R(\vec{s})|-\Sigma)=\Pr(\vec{s}|\Sigma).
\end{align*}
Thus, the entropy conditioned current is an odd function of the entropy:
\begin{align*}
j(-\Sigma)& = \sum_{J} J \Pr(J|-\Sigma)
\\ & = \sum_{J,\vec{s}} J \Pr(J,\vec{s}|-\Sigma)
\\ & = \sum_{J,\vec{s}} J \delta_{J,J(\vec{s})}\Pr(\vec{s}|-\Sigma)
\\ & = \sum_{\vec{s}}J(\vec{s})\Pr(\vec{s}|-\Sigma)
\\ & =  \sum_{\vec{s}}J(R(\vec{s})) \Pr(R(\vec{s})|-\Sigma)
\\ & =  \sum_{\vec{s}}-J(\vec{s}) \Pr(\vec{s}|\Sigma)
\\ &=  -j(\Sigma).
\end{align*}

Having assumed that $\Sigma$ is a function of the state trajectory, we can re-express the TSSC fluctuation theorem:
\begin{align*}
\Pr(R(\vec{s}))\delta_{\Sigma(R(\vec{s})),-\Sigma'}=e^{-\Sigma'}\Pr(\vec{s})\delta_{\Sigma', \Sigma(\vec{s})},
\end{align*}
This can only be true if the entropy production is itself a current $\Sigma(R(\vec{s}))=-\Sigma(\vec{s})$.  Thus, we see that our new function of the trajectories $J'$ is indeed a current as well:
\begin{align*}
    J'(R(\vec{s}))& = j(\Sigma(R(\vec{s})))
    \\& = j(-\Sigma(\vec{s}))
    \\& =- j(\Sigma(\vec{s}))
    \\ & =-J'(\vec{s}).
\end{align*}

\subsection{Minimizing Current}

Given that any current's scaled variance can be reduced by finding its corresponding entropy-conditioned current, given some TSCC process $\Pr(\vec{s}, \Sigma)$, we need only find the function $j(\Sigma)$ that minimizes the scaled variance $\epsilon^2_{j}=\frac{\langle j^2\rangle }{\langle j \rangle^2}-1$.  There is a single important constraint that applies to these functions, which is that $j(-\Sigma)=-j(\Sigma)$, meaning that this is a constrained optimization.  

However, we can ignore this constraint by using the TSCC DFT and summing over the state trajectories of Eq. \ref{eq:NESSFT} to produce a familiar relation \cite{Hase19a,Timp19a}:
\begin{align}
\Pr(-\Sigma)=e^{-\Sigma}\Pr(\Sigma).
\label{appeq:NESSFT}
\end{align}
We use this to express the average $j$ and $j^2$ can in terms of positive entropy productions:
\begin{align}
\langle j^2 \rangle & = \sum_{\Sigma > 0} \Pr(\Sigma)(1+e^{-\Sigma})j(\Sigma)^2
\\ \langle j \rangle & = \sum_{\Sigma > 0} \Pr(\Sigma)(1-e^{-\Sigma})j(\Sigma). \nonumber
\label{appeq:averages}
\end{align}
This means that $\epsilon^2_j$ can be expressed in terms of only positive entropy productions.  $j(\Sigma)$ is \emph{unconstrained} over positive $\Sigma$, so the minimum occurs when:
\begin{align}
\frac{\partial}{\partial j(\Sigma)} \epsilon^2_j & =  \frac{1}{\langle j \rangle^2}\left( \frac{\partial \langle j^2\rangle } {\partial j(\Sigma)} - \frac{2 \langle j^2 \rangle}{\langle j \rangle} \frac{ \langle j \rangle } {\partial j(\Sigma)} \right) 
\\ &  =0 \text{ for all } \Sigma > 0. \nonumber
\label{appeq:Optimum}
\end{align}
Applying the derivative with respect to the positive-entropy current $j(\Sigma)$ to the averages shown in Eq. (\ref{appeq:averages}) yields:
\begin{align*}
\frac{\partial \langle j \rangle }{\partial j(\Sigma)} & =(1-\epsilon^{-\Sigma})\Pr(\Sigma)
\\\frac{\partial \langle j^2 \rangle }{\partial j(\Sigma)} & =(1+\epsilon^{-\Sigma})\Pr(\Sigma)2j(\Sigma).
\end{align*}
Finally, plugging these into Eq. (\ref{appeq:averages}), we solve for the current with the minimum scaled variance:
\begin{align}
\boxed{j_\text{min}(\Sigma)  = \frac{\langle j^2_\text{min} \rangle}{\langle j_\text{min} \rangle} \frac{1-e^{-\Sigma}}{1+e^{-\Sigma}}}\,,
\end{align}
which applies as long as $\Sigma$ is in the support of the entropy distribution.  Note that, even though the expression for the minimal current was derived for $\Sigma>0$, it applies to $\Sigma \leq 0$ as well, because of the condition that a current must satisfy $j(-\Sigma)=-j(\Sigma)$.  Indeed, $j_\text{min}(0)=0$ and $j_\text{min}(-\Sigma)=\frac{\langle j^2 \rangle}{\langle j \rangle} \frac{1-e^{\Sigma}}{1+e^{\Sigma}}=\frac{\langle j^2 \rangle}{\langle j \rangle} \frac{e^{-\Sigma}-1}{e^{\Sigma}+1}=-j_\text{min}(\Sigma)$.  Thus, we have found the form of the current that minimizes scaled variance, and it depends exclusively on the entropy production of the process:
\begin{align*}
J_\text{min}(\vec{s})= j_\text{min}(\Sigma(\vec{s})).
\end{align*}
This result can also be found by recognizing that the bound described in Ref.~\cite{falasco2020unifying} is achieved when the current is proportional to their averaging observable.

This relation can be inverted as well.  If we manage to discover a maximum-precision current $J_\text{min}(\vec{s})$ for a thermodynamic process, then the entropy production can be exactly calculated as a function of the trajectory:
\begin{align}
\boxed{\Sigma(\vec{s})  = \ln \frac{\langle J^2_\text{min} \rangle + \langle J_\text{min} \rangle J_\text{min}(\vec{s})}{\langle J^2_\text{min} \rangle - \langle J_\text{min} \rangle J_\text{min}(\vec{s})} }.
\end{align}
This provides a powerful tool to infer not just the average entropy production, but the entire distribution of entropy productions.

\subsection{Thermdodynamic Bound on Scaled Variance}

For simplicity, note that we can choose any real value for $k=\langle j^2_\text{min} \rangle/\langle j_\text{min} \rangle$, and the entropy-conditioned current $j_\text{min}(\Sigma)$ will minimize the scaled variance.  Moreover, $\frac{1-e^{-\Sigma}}{1+e^{-\Sigma}}=\tanh(\Sigma/2)$.  Whatever the TSCC entropy production distribution $\Pr(\Sigma)$ may be, the current:
\begin{align}
J_\text{min}(\vec{s})=k \tanh(\Sigma(\vec{s})/2),
\end{align}
will set a lower bound on the all other currents of the system:
\begin{align}
\epsilon^2_J & \geq \epsilon^2_{J\text{min}}
\\ & = \frac{\langle \tanh(\Sigma/2)^2 \rangle }{\langle \tanh(\Sigma/2) \rangle^2}-1, \nonumber
\end{align}
where the constant $k=\langle j^2 \rangle/\langle j \rangle$ has factored out.  This bound on the scaled variance is \emph{tight}, because it is realized by our newly defined $J_\text{min}(\vec{s})$. For this reason, $\epsilon^2_{J_\text{min}}$ represents the tightest possible bound on the scaled variance for the TSCC process $\Pr(\Sigma, \vec{s})$.

Once again, the TSCC DFT ($\Pr(\Sigma)=e^{\Sigma}\Pr(-\Sigma)$) simplifies:
\begin{align}
\label{appeq:TanhIdentity}
\langle \tanh(\Sigma/2)^2 \rangle  &  =  \sum_{\Sigma>0} \Pr(\Sigma)(1+e^{-\Sigma}) \left(\frac{1-e^{-\Sigma}}{1+e^{-\Sigma}} \right)^2 \nonumber
\\ & =  \sum_{\Sigma>0} \Pr(\Sigma)(1-e^{-\Sigma}) \left(\frac{1-e^{-\Sigma}}{1+e^{-\Sigma}} \right) \nonumber
\\ & = \langle \tanh(\Sigma/2) \rangle.
\end{align}
As a result, we have the simplified bound on the scaled variance in terms of the entropy production:
\begin{align}
\label{appeq:EMin}
\boxed{ \epsilon^2_J  \geq \epsilon^2_{J\text{min}}= \frac{1}{\langle \tanh(\Sigma/2) \rangle}-1}.
\end{align}
This echos past thermodynamic uncertainty relations in that it relates the minimum variance to the entropy production.  

However, unlike $\epsilon^2_{HG}$, $\epsilon^2_{HVV}$, and $\epsilon^2_{TGGL}$, which are all functions of the average entropy production $\langle \Sigma \rangle$, this bound depends on an \emph{average of a function of the entropy}.  This means that higher moments of the entropy distribution will appear in the bound, not just the average entropy production $\langle \Sigma \rangle$.  Specifically, if $f(x)=\tanh (x/2)$, then we can express the bound in terms of higher moments:
\begin{align}\label{appeq:EminMoments}
\epsilon^2_{J\text{min}}=\frac{1}{ f(\langle \Sigma \rangle) +\sum_{n=2}^\infty \frac{f^{(n)}(\langle \Sigma \rangle)}{n!} \langle (\Sigma-\langle \Sigma \rangle)^n\rangle }-1.
\end{align}
If we ignore the higher moments, $\epsilon^2_{J\text{min}}$ simplifies to $\frac{1}{\langle \tanh (\Sigma/2) \rangle}-1=\frac{2}{e^{\langle \Sigma \rangle} -1} = \epsilon^2_\text{HVV}$.  Because $\epsilon^2_{J\text{min}}>\epsilon^2_\text{HVV}$, the higher momentum terms are negative $\sum_{n=2}^\infty \frac{f^{(n)}(\langle \Sigma \rangle)}{n!} \langle (\Sigma-\langle \Sigma \rangle)^n\rangle<0$, and thus raise the minimum scaled-variance.

\section{Background}
\label{app:Background}

Barato and Seifert \cite{Bara15a} derived the first thermodynamic uncertainty relation in terms of rates of entropy production. They considered the precision of currents $J$ through the inverse of the signal to noise ratio: the scaled variance:
\begin{align}
\epsilon^2_J \equiv \frac{\text{var}(J)}{\langle J \rangle^2}.
\end{align}
Remarkably, they found that the precision couldn't be maximized without a corresponding increase in the average entropy production:
\begin{align}
\epsilon^2_J \geq \frac{2}{\langle \Sigma \rangle}.
\end{align}
Because Barato and Seifert discovered this form of the bound, we denote it the Barato-Seifert (BS) bound on scaled variance:
\begin{align}
\epsilon^2_\text{BS} \equiv \frac{2}{\langle \Sigma \rangle}.
\end{align}

Further exploration found that detailed fluctuation theorems \cite{Jarz00a,Croo99a} can be used to prove modified thermodynamic uncertainty relations.  The system $\mathcal{S}$ may be influenced by an external control parameter $\lambda_t$ at every time $t$, thereby performing a computation over the time interval $t \in (0,\tau)$. Under time-symmetric control $\lambda^\dagger_{\tau-t}=\lambda_t$ and conjugation of the distribution under the operation $\Pr(s,\tau)=\Pr(s^\dagger,0)$ \footnote{Reference \cite{Hase19a} makes a slightly different assumption, which is that the computation preserves the initial distribution $\Pr(s,\tau)=\Pr(s,0)$.  Their assumption is often sufficient, when time-antisymmetric variables like momentum are thermalized, such as in overdamped Langevin dynamics.  However, it is necessary to account for the conjugation of the system state to establish the most general conditions for the DFT.} the probability of a reverse trajectory is exponentially damped by the entropy production \cite{Hase19a}:
\begin{align}
\Pr(R(\vec{s}),-\Sigma)=e^{-\Sigma}\Pr(\vec{s},\Sigma).
\label{app:NESSFT}
\end{align}
This is the Time-Symmetrically Controlled Computation (TSCC) Detailed Fluctuation Theorem (DFT), whose derivation we detail in Ref. \cite{Supp22a} \ref{app:TSCC DFT}.  This version of the DFT includes NESS dynamics, for which the control parameter is constant $\lambda_t=\lambda_{t'}$, as it expresses the relation originally derived by Evans and Searles \cite{Evan02a}.  It can also  describe, as we explore here, computations that begin in equilibrium, and are then allowed to relax after the application of time-symmetric control signal.  Such symmetries are ubiquitous in computing \cite{Riec19b}.

Hasegawa and Van Vu \cite{Hase19a} used Eq. (\ref{app:NESSFT}) to demonstrate that the scaled variance is bounded below by:
\begin{align}
\epsilon^2_J \geq \frac{2}{e^{\langle \Sigma \rangle}-1} \equiv \epsilon^2_\text{HVV},
\end{align}
where we have labeled their bound by $\epsilon^2_\text{HVV}$.  Also using the fluctuation theorem, Timparano, Guarnieri, Goold, and Landi \cite{Timp19a} showed another bound using the average entropy production:
\begin{align}
\epsilon_J^2 \geq  \text{csch}^2[g(\langle \Sigma \rangle)] \equiv \epsilon^2_\text{TGGL}
  ~,
\end{align}
where $g(x)$ is the inverse of $x\tanh (x)$, and we have again labeled the bound for the authors.  This bound is tighter than $\epsilon^2_\text{HVV}$. In fact, it is the tightest possible bound on scaled variance that can be determined from the average entropy \cite{Timp19a}.

All three bounds above, while far from complete list, are functions of average entropy production, so we can consider which set the tightest and most accurate bounds for different TSCCs when taking into account the effects of higher moments. 

It is also worth noting that our results is reminiscent of the recently introduced notion of \textit{hyper-accurate currents}~\cite{Busiello2019,Andre2021}, defined as those possessing the maximum signal-to-noise ratio. While in that case, however, the form of these currents, whose precision therefore can be used to bound the precision of any other thermodynamic current (thus much alike, in spirit, to Eq.~\eqref{eq:EMin}) was found within classical and quantum thermoelectrics given a coherent transport modelization in the Landauer-B\"uttiker formalism, in our work we derive them by imposing the TSCC symmetry on $\Pr(\vec{s},\Sigma)$.

\section{Proof that $\epsilon^2_J \geq \emin \geq \TGGL $}
\label{app:BoundOrdering}
\vspace{1cm}

Let us start by considering the expressions for the two TUR bounds $\epsilon^2_{j_{min}}$ and $\epsilon^2_{TGGL}$, which can be proven to bound the noise-to-signal ratio (or scaled variance) $ \epsilon^2_J \equiv \frac{\mathrm{Var}(J)}{\langle J \rangle^2} $ of any current $J$ (assumed to be anti-symmetric under time-reversal) in steady-state regime under the constraint that the probability distribution for the entropy production $\Sigma$ satisfies the fluctuation relation symmetry:
\begin{equation}\label{appeq:FR}
    \Pr(-\Sigma) = \Pr(\Sigma) e^{-\Sigma}.
\end{equation}
Our main result is that, under no additional assumptions about the distribution over $\Sigma$, the following bound can be placed:
\begin{equation}\label{apeq:emin}
    \epsilon^2_J \geq \epsilon^2_{j_{min}} \equiv \frac{1}{\langle \tanh(\Sigma/2)\rangle} -1.
\end{equation}
The other bound, proven in Ref. ~\cite{Timp19a} shows instead if on top of the above the averages of the entropy production and of a generic current, i.e., $\langle \Sigma\rangle$ and $\langle J\rangle$, are fixed and satisfy \textit{a joint} fluctuation relation symmetry of the form:
\begin{equation}
    \Pr(-\Sigma,J) = \Pr(\Sigma,J) e^{-\Sigma},
\end{equation}
then the following TUR bound can be derived:
\begin{equation}\label{appeq:TURdeForce}
    \epsilon^2_J \geq \epsilon^2_{TGGL} \equiv \text{csch}^2(g(\langle \Sigma\rangle/2)).
\end{equation}
where $g(\langle \Sigma\rangle)$ is the function inverse of $\langle \Sigma\rangle\tanh(\langle \Sigma\rangle)$.

Is there a relationship between these two bounds? In these brief notes we provide a positive answer to this question.

The first step to show this is to re-express the right hand side of Eq.~\eqref{appeq:TURdeForce} as:
\begin{equation}
    \epsilon^2_J \geq \epsilon^2_{TGGL} \equiv \frac{1}{\tanh^2(g(\langle \Sigma\rangle/2))}-1
\end{equation}
by using the identity $\text{csch}^2(x) + 1 = 1/\tanh^2(x)$.
In order to proceed, let us then define the following quantities:
\begin{align}
z(\Sigma) \equiv \tanh^2\left(\frac{\Sigma}{2}\right)\\
h(\Sigma) \equiv \Sigma \tanh\left(\frac{\Sigma}{2}\right).
\end{align}
Furthermore, let us introduce the inverse function of $h(\Sigma)$ and denote with by $g$, i.e. $g(h(\Sigma)) = \Sigma$. Notice that this is possible since $h(\Sigma)$ is a monotonically increasing function of $\Sigma$ whenever $\Sigma \geq 0$.
This is not a limitation, since, thanks to the fluctuation relation symmetry Eq.~\eqref{appeq:FR}, it amounts to consider:
\begin{equation}
  \langle(\ldots)\rangle \equiv \sum_{\Sigma} (\ldots) \Pr(\Sigma) = \sum_{\Sigma>0} (\ldots) \Pr(\Sigma)\left(1 + e^{-\Sigma}\right).
\end{equation}
The way to make use of the above-introduced quantities is to realize that the composite function 
$w(h) \equiv f(g(h))$ is a concave function of $h$ (when $h\geq 0$), since $w'(h) > 0$ $w''(h) < 0$. 
This allows us to exploit Jensen's inequality (with the ``norm'' being given by the average $\langle\ldots\rangle$ calculated with the probability distribution $\Pr(\Sigma)(1+e^{-\Sigma})$, as discussed a few lines ago) and obtain the following inequality:
\begin{align}\label{appeq:Inequality}
\langle \tanh\left(\frac{\Sigma}{2}\right)^2 \rangle &= \langle z(\Sigma) \rangle = \langle z(g(h(\Sigma))) \rangle \\
& \leq z(g(\langle h(\Sigma) \rangle )) = z(g(\langle \Sigma \rangle)),
\end{align}
where on the last passage we have used the fact that $\langle \Sigma \rangle = \langle h(\Sigma) \rangle$. Notice that the quantity that appears on the right hand side of this inequality, i.e. 
$z(g(\langle \Sigma \rangle))$, is nothing but the denominator in the first term of the r.h.s. of Eq.~\eqref{appeq:TURdeForce}.

The final step to conclude the proof then consists in realizing that Eq. \ref{appeq:TanhIdentity} means that the the l.h.s. of the inequality Eq.~\eqref{appeq:Inequality} is the denominator in the first term of the r.h.s. of Eq.~\eqref{appeq:EMin}.

In view of Eq.~\eqref{appeq:Inequality} then immediately follows that:
\begin{equation}
    \epsilon^2_{j_{min}} \geq \epsilon^2_{TGGL}.
\end{equation}

\section{Simulations}
\label{app:Sim}

\begin{figure}
    \centering
    \subfigure(a){\includegraphics[width=.28\columnwidth]{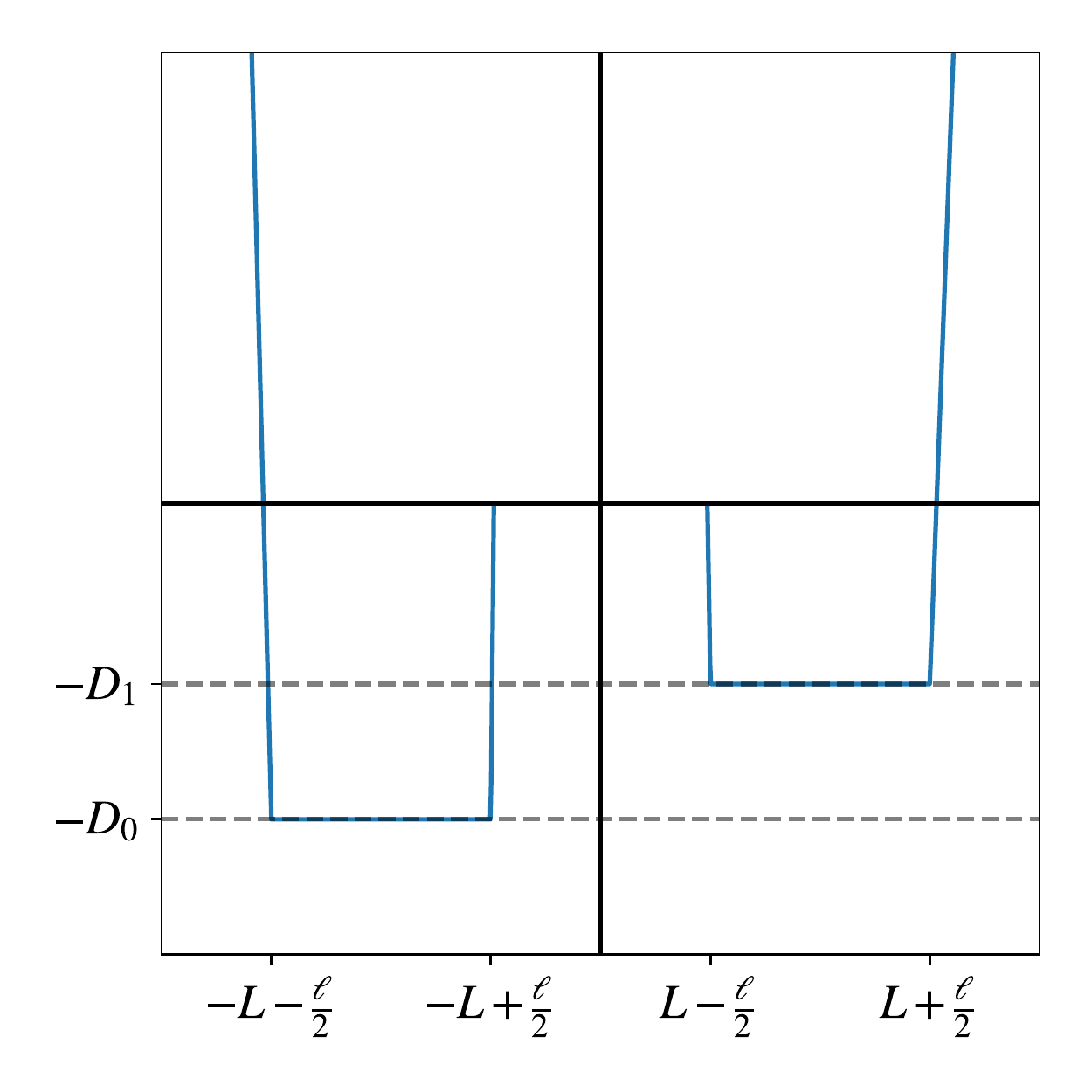}} 
    \subfigure(b){\includegraphics[width=.28\columnwidth]{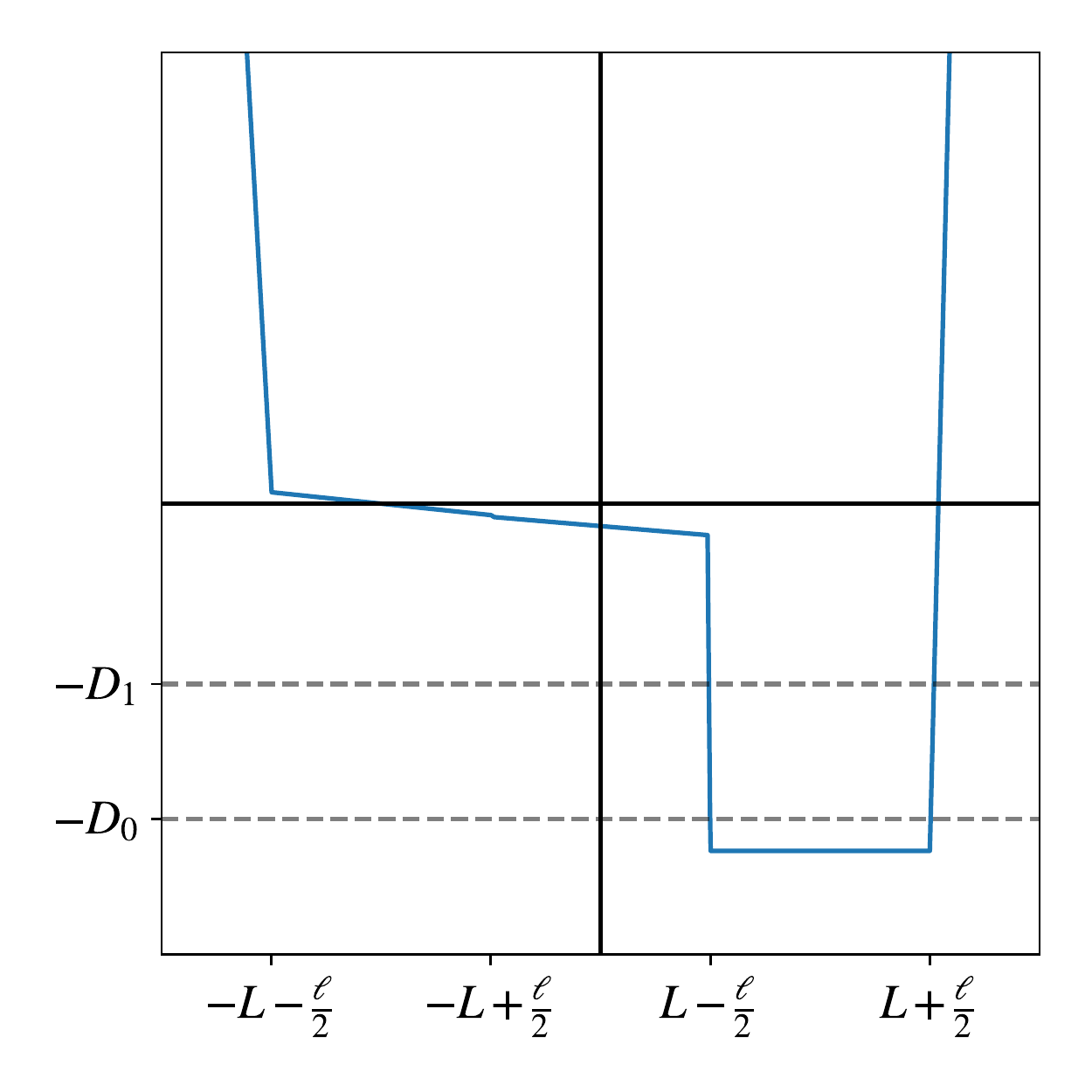}} 
    \subfigure(c){\includegraphics[width=.28\columnwidth]{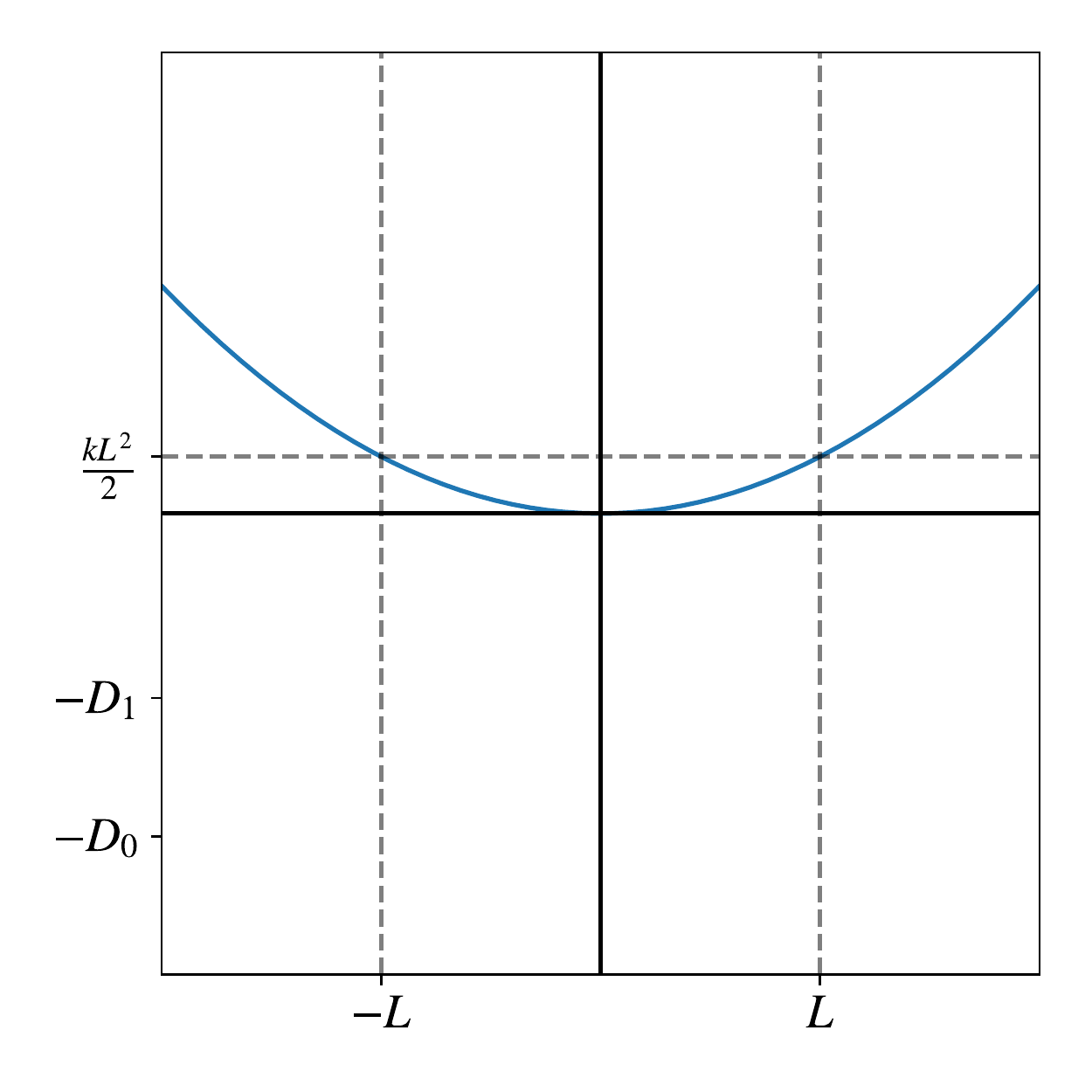}}
    \caption{Potential energy landscapes during (a) storage ($\US$), (b) the reset computation, and (c) the swap computation. The offset from $D_0$ in the right well during the reset computation creates the same energy barrier as the left well during the storage potential.}
    \label{fig:Potentials}
\end{figure}

All of the TSCC protocols used to generate Fig. \ref{fig:JminSim} follow the same qualitative structure. The system begins in equilibrium with a thermal reservoir, exposed to a storage potential $\US$. At $t=0$, a computational potential $\UC$ is applied until $t=\tau$ and then the system is re-exposed to $\US$. Because the potential energy surface changes only at $t=0$ and $t=\tau$, the work $W(x_0, x_\tau, \UC, \US)$ done during any particular microscopic trajectory is given by the sum of the work invested at $t=0$ and that invested at $t=\tau$. Once the system relaxes back to equilibrium, any energy added to the system in the form of work has been dissipated into the environment: generating entropy in the heat bath equal to $\beta W$. Thus, the entropy generated by a trajectory can be calculated simply as:
\begin{align}
\label{appeq:SigmaCalc}
\beta^{-1} \Sigma(\vec{x}) = \beta^{-1} \Sigma(x_0,x_\tau)  = W_0 &+ W_\tau \nonumber \\
= \UC(x_0)-\US(x_0) &+ \US(x_\tau)-\UC(x_\tau)
\end{align}
A large enough ensemble of initial conditions allows for an estimate of the entropy production distribution and, through Eq. (\ref{eq:EMin}), an estimate of the minimum scaled variance that can be achieved by any current defined on the system. 
For both the Reset and the Swap operations $\US$ was chosen to be an asymmetric double square-well potential with wells of depths $D_0, D_1$, widths $\ell$, and centered at $x=\pm L$ (See Fig. \ref{fig:Potentials}.)

\subsection{Reset}
\label{app:Reset}
The reset simulation used nondimensionalized overdamped Lanvegin dynamics:
\begin{align*}
  dx &= - \Omega \partial_{x} U(x,t) dt
   + \xi \sqrt{2}\, r(t) \sqrt{dt}
  ~,
\end{align*}
Here, $r(t)$ is a memoryless Gaussian variable, and all parameters and variables have been scaled to be dimensionless by the scheme $q' =q \cdot q_c $. $q'$ is some dimensional quantity with $q_c$ a scaling factor and $q$ the dimensionless variable. The dimensionless simulation parameters $\Omega$ and $\xi$ are combinations of the scaling factors and the familiar dimensional Langevin parameters. For all overdamped simulations, $\Omega = \xi = 1$. This represents some relationship between the physical parameters of the system, but the exact relationship is not important for our purposes. Note also that we choose our scaling factor for energies to be $\kB T$ so that the potential energy can be thought of as being in units of the thermal energy.

The computational potential used for the reset is shown in Fig. \ref{fig:Potentials}, and was held for $\tau=2$ time units. The left well is turned into a ramp leading in the right one, and the right wells energy is lowered. This causes nearly all realizations of the process to fall into the right well, hence the name ``reset''. As the wells get narrower and narrower, the variance of different work values conditioned on being in the left or right at $t=0$ will shrink---however, this distribution can never be a bimodal distribution due to the necessary presence of $0$ entropy production events. As such, we expect the minimum scaled variance of any current from these simulations to follow the ``discrete bound'' derived in following section.

To show that this class of protocol obeys the bound, a suite of \resetN simulations was performed using a Monte Carlo Markov Chain (MCMC) inspired approach to find parameters for which $\epsilon^2_{J\text{min}}$, as estimated by Eq. (\ref{eq:EMin}), is minimized. On each iteration of algorithm, a new value was chosen for 2 (chosen randomly, with replacement) of the 4 parameters $L,\ell, D_0, D_1$ using a Gaussian distribution centered on its current value, checking to make sure that $\ell<L$ and $D_0 > D_1$. After performing the simulation and measuring $\emin$, the proposed parameter change was accepted with certainty if the new $\emin$ was less than the original and accepted with a probability $p \propto e^{-\Delta \emin}$ if it was greater. Jumps for which the average entropy production did not satisfy $1.5 \leq \langle \beta\Sigma \rangle \leq 6$ we also rejected, to keep the algorithm from exploring an untenable range of parameter space. The end result is that for the simulations in Fig. \ref{fig:JminSim} the parameters were sampled from the following ranges, though not uniformly or independently: $L \in (.2,1.2)$, $\ell \in (0,1.1)$, $D_0 \in (1,6.2)$ and $D_1 \in (.2,3.6)$. Here, $L,\ell$ are in units of the non-dimensional position and $D_0, D_1$ in units of $\kB T$.

\subsubsection{Discrete Bound}
\label{app:DiscreteBound}

To see why the reset operation as described cannot generate a truly bimodal distribution, consider a simplified version of the continuous state dynamics that implement the reset operation: a two-level system operating in the regime of rate equation dynamics. Here, the ``potential energy landscape'' is defined simply by setting the energy levels of the two states $x\in \{A,B\}$. The $\US$ energy levels are $E_A = E, E_B =0$ so that the equilibrium distribution over the two states is given by $\rho_0 = (Pr(X_0=A), Pr(X_0=B)) =  (p_E, 1-p_E)$. Here, $\UC$ will swap the two energy levels so that  $E_A = 0, E_B = E$ and $\tau$ will be long enough that the system has enough time to equilibrate to $\UC$ yielding $\rho_\tau = (1-p_E,p_E)$. Using Eq. (\ref{appeq:SigmaCalc}) reveals only three possible outcomes for $\Sigma(x_0,x_\tau)$:
\begin{align*}
\Sigma(A,A) = \Sigma(B,B) = 0 \\
\Sigma(B,A) = -\Sigma(A,B) = 2\beta E.
\end{align*}
Since the system has been given time to equilibrate the state at time $t=0$ is not correlated with the state at time $t=\tau$ so the probabilities of these different events can be readily calculated, yielding the full distribution of entropy production:
\begin{align*}
Pr(\Sigma(A,A)) = Pr(\Sigma(B,B))= p_E*(1-p_E) \\
Pr(\Sigma(A,B)) = p_E^2 \\
Pr(\Sigma(B,A)) = (1-p_E)^2 \\
\end{align*}
For the two-level system, $p_E = \frac{e^{-\beta E}}{1+e^{-\beta E}}$. For any anti-symmetric function $J(\Sigma)=-J(-\Sigma)$, we have:
\begin{align*}
 \langle J(\Sigma) \rangle (E) &= J(2 \beta E) ( (1-p_E)^2 - p_E^2) \\
 &= J(2\beta E) ( 1-2p_E) = J(2\beta E) \frac{1-e^{-\beta E}}{1+e^{-\beta E}}\\
 & = J(2\beta E) \tanh ( \beta E /2).
\end{align*}
The zero entropy events do not appear directly in the first line because $J(0)=0$ for any function $J$ that is odd in $\Sigma$. We use this equation to readily calculate both the average entropy production:
\begin{align}
\label{appeq:DiscreteAverage}
 \langle \Sigma \rangle (E) &= 2\beta E \tanh ( \beta E /2),
\end{align}
and the minimum variance current for the distribution (through Eq. (\ref{eq:EMin})):
\begin{align}\label{appeq:DiscreteJmin}
 \emin(E) &= \frac{1}{\langle \tanh(\Sigma/2) \rangle} -1 \nonumber \\
 &= \frac{1}{\tanh(\beta E) \tanh ( \beta E /2)}.
\end{align}
We can then use the parameter $E$ to find the effective bound that Eq. (\ref{eq:JMin}) sets for a given average entropy production in the reset process. Figure \ref{fig:JminSim} shows that, while this bound lies below the one for a normally distributed entropy production---it lies far above previous bounds, $\TGGL$ and $\HVV$, that used only the TSCC DFT in Eq. (\ref{eq:NESSFT}). And, for the reset processes simulations the bound appears tight. This example showcases the flexibility of Eq. (\ref{eq:JMin}), as we see it can be used to set operationally useful regime and/or protocol specific bounds by including information about the system of interest.

\subsection{Swap}
\label{app:Swap}

% parameters changed: location, localization, d_0, d_1, lambda
The swap simulations used non-dimensionalized underdamped Langevin dynamics:
\begin{align*}
  dx &= v dt \\
  dv &= -\lambda v dt - \Theta \partial_{x} U(x,t) dt
   + \eta \sqrt{2\lambda}\, r(t) \sqrt{dt}
  ~,
\end{align*}

A similar scaling strategy as that described in the previous section leads to three dimensionless parameters: $\lambda, \Theta, \eta$. $\Theta=\eta=1$ for all simulations but $\lambda$, which parameterizes the system's coupling to its thermal environment, was allowed to change.

The computational potential for the swap is a harmonic potential with $k=m\pi^2$ (see Fig. \ref{fig:Potentials}.) If $\lambda=0$, the system undergoes a harmonic oscillation with a period of $2$ time units. Exactly halfway through the oscillation, the particles that were in the left(right) well of $\US$ should now be located where the right(left) well is. Thus, if $\US$ is turned back on at $t=1$ we have implemented a `swap' operation between asymmetric wells. Because the dynamics are underdamped, this type of protocol also persists in the case of nonzero $\lambda$: the system will undergo the same oscillation approximately, with some amount of dissipation and stochasticity. The work cost distribution to implement this protocol will approach a bimodal distribution, with particles starting in the left well costing an energy value near $D_0-D_1$ and those starting in the right yielding an energy surplus near $D_1-D_0$. This distribution is sharpened by narrower wells and lower values of $\lambda$.

Instead of attempting to minimize $\emin$, the point of this simulation is to showcase that $\emin$ faithfully captures all cases: where from where $\TGGL$ is tight, to where $\BS$ is tight, to where neither is a good approximation. The protocol described above generates the bimodal distribution that saturates $\TGGL$ under some cases, but can also produce entropy production distributions where the minimal scaled variance is well above $\BS$. To showcase this variety, a MCMC approach was again used. On each iteration of algorithm, a new value was chosen for 3 (chosen randomly, with replacement) of the 5 parameters $L,\ell, D_0, D_1, \lambda$ using a Gaussian distribution centered on its current value, checking to make sure that $\ell<L$, $D_0 > D_1$ ,and $\lambda>0$. In this case, all jumps for which $\langle \Sigma \rangle$ fell between $2 $ and $5 $ were accepted, and those that did not were accepted with a probability that exponentially decayed in $|\langle \Sigma \rangle - 3.5| $. The plot in Fig. \ref{fig:JminSim} shows a suite of \swapN simulations, that stem from 8 different starting points for $\lambda \in (0,.15)$, but all other parameters the same. As the algorithm evolved, all free parameters were allowed to shift with the result being that paramters were sampled from the following ranges: $L \in (.14,.68)$, $\ell \in (0,.39)$, $D_0 \in (1.2,10.5)$ and $D_1 \in (.39,3.6)$, $\lambda \in (0,.2)$.

\subsection{Simulation Details}

\def \dt {$5\times10^{-5}$}
\def \Nsamples {50,000}

\begin{figure}
    \centering
    \includegraphics[width=.5\columnwidth]{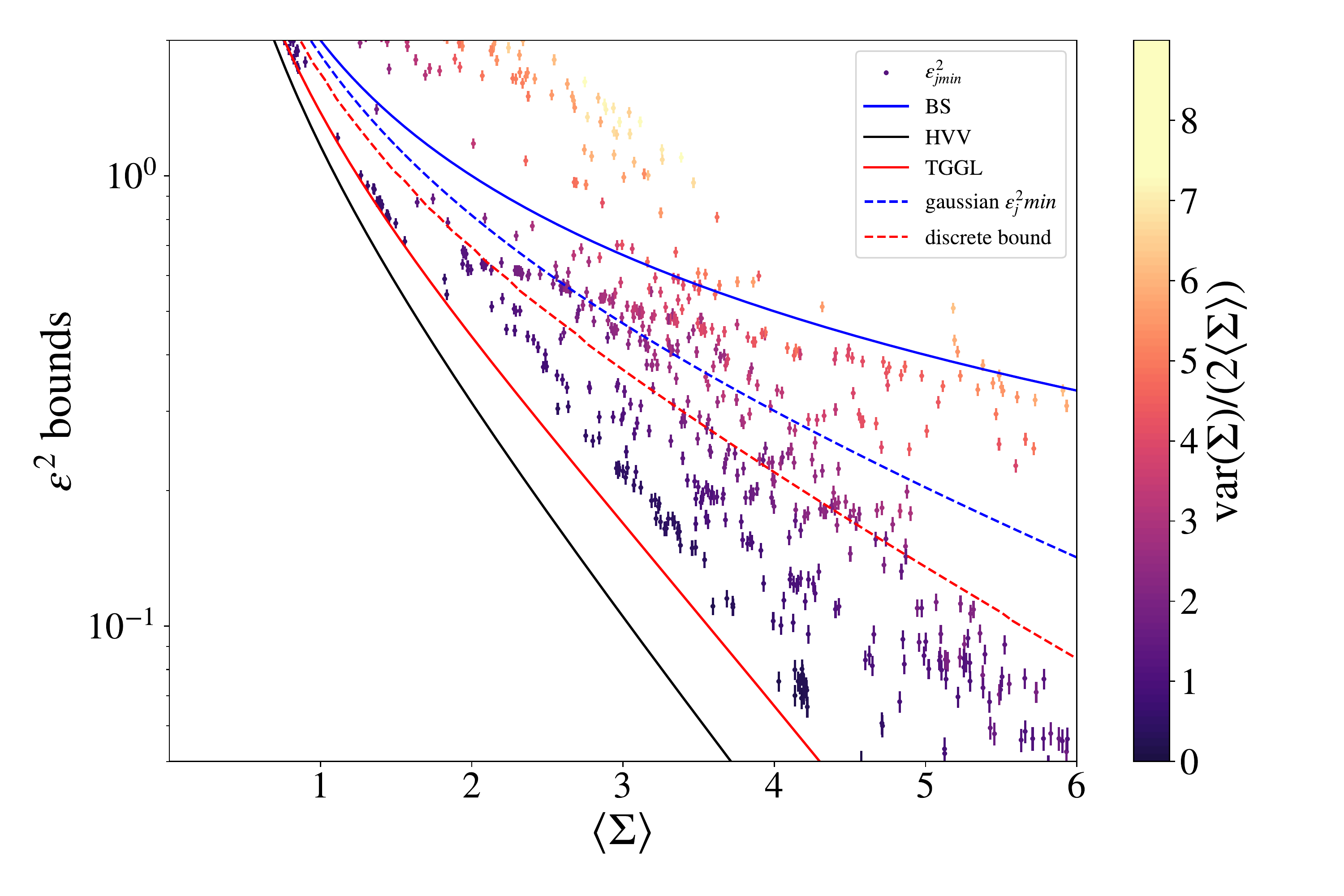}
    \caption{Same simulation data that generated Fig. \ref{fig:JminSim}, but with $3\Sigma$ error bars calculated from the full simulation, rather than just the well-sampled positive events.}
    \label{fig:ErrorbarExample}
\end{figure}

Each dot in the simulation plot was calculated from an ensemble of $\Nsamples$ trajectories sampled from the equilibrium distribution, using a Monte Carlo method. To reduce the size of error bars, averages shown in Fig. \ref{fig:JminSim} were calculated using only the positive entropy production events according to Eq. (\ref{appeq:averages}) that assumes a system that obeys the TSCC DFT. This numerical trick does not change any of the qualitative results, but allows for plots in which the error bars are small enough to not be relevant in the plot. For example, Fig. \ref{fig:ErrorbarExample} shows a plot of the same simulation data that generated Fig. \ref{fig:JminSim}, but with $3\Sigma$ error bars calculated from the full simulation, rather than just the well-sampled positive events. The Langevin simulations of the dimensionless equations of motion for both the underdamped and overdamped cases employed a fourth-order Runge-Kutta method for the deterministic portion and Euler's method for the stochastic portion of the integration with $dt$ set to \dt. Python NumPy's Gaussian number generator was used to generate the memoryless Gaussian variable r(t).

\end{document}

%% file: cmechabbrev.tex
\newcommand{\CausalState}   { \mathcal{S} }

\newcommand{\forward}{+}
\newcommand{\reverse}{-}
\newcommand{\forwardreverse}{\pm} % \pm

\newcommand{\FutureCausalState} { {\CausalState}^{\forward} }

\newcommand{\PastCausalState}   { {\CausalState}^{\reverse} }

\newcommand{\lastindex}[2]{
  \edef\tempa{0}
  \edef\tempb{#2}
  \ifx\tempa\tempb
    \edef\tempc{#1}
  \else
    \edef\tempa{0}
    \edef\tempb{#1}
    \ifx\tempa\tempb
      \edef\tempc{#2}
    \else
      \edef\tempc{#1+#2}
    \fi
  \fi
  \tempc
}

\newcommand{\CSjoint}[1][,]{
   \edef\tempa{:}
   \edef\tempb{#1}
   \ifx\tempa\tempb
      \ensuremath{\FutureCausalState\!#1\PastCausalState}
   \else
      \ensuremath{\FutureCausalState#1\PastCausalState}
   \fi
}

\newif\ifpm
\edef\tempa{\forwardreverse}
\edef\tempb{\pm}
\ifx\tempa\tempb
   \pmfalse
\else
   \pmtrue
\fi